\documentclass[a4paper,11pt]{article}
\usepackage{jheppub}

\pdfoutput=1
\usepackage{hyperref}
\usepackage{graphicx}
\usepackage[utf8]{inputenc}
\usepackage{xspace}
\usepackage{amsmath,amssymb}
\usepackage{bbm}
\usepackage{xcolor}
\usepackage{booktabs}
\usepackage{subcaption}
\usepackage{mathtools}
\usepackage[ruled,norelsize]{algorithm2e}
\usepackage{empheq}
\usepackage{cancel}
\usepackage{multirow}
\usepackage{slashed}
\usepackage{feynmp}
\usepackage{feynmp-auto}
\usepackage{tikz}
\usetikzlibrary{decorations.pathmorphing}
\usetikzlibrary{decorations.markings}
\usetikzlibrary{positioning, shapes, snakes, arrows}
\tikzset{
	quark/.style={postaction={decorate},
		decoration={markings,mark=at position .5 with {\arrow[#1]{latex}}}},
	scalar/.style={dashed,postaction={decorate},
		decoration={markings,mark=at position .5 with {\arrow[#1]{latex}}}},
	gluon/.style={decorate,
		decoration={coil,amplitude=2pt, segment length=2pt,  pre length=.1cm, post length=.1cm}},
	boson/.style={-latex,decorate, decoration={snake, segment length=4pt, amplitude=1.8pt, pre length=.1cm, post length=.25cm}},
	photon/.style={decorate, decoration={snake, segment length=4pt, amplitude=1.8pt,  pre length=.1cm, post length=.1cm}},
	dphoton/.style={decorate, decoration={snake, segment length=4pt, amplitude=1.8pt,  pre length=.1cm, post length=.25cm},-latex}
}

\newcommand{\sd}{d}

\newcommand{\MSbar}{\ensuremath{\overline{\text{MS}}}\xspace}

\newcounter{BEQ}

\DeclareFontFamily{U}{rcjhbltx}{}
\DeclareFontShape{U}{rcjhbltx}{m}{n}{<->rcjhbltx}{}

\title{Two-loop anomalous dimensions for small-$R$ jet versus hadronic
  fragmentation functions}

\author[a]{Melissa van Beekveld,}
\author[b]{Mrinal Dasgupta,}
\author[b, c]{Basem Kamal El-Menoufi,}
\author[d]{Jack Helliwell,}
\author[e]{Alexander Karlberg,}
\author[e]{Pier Francesco Monni}

\affiliation[a]{Nikhef, Theory Group, Science Park 105, 1098 XG, Amsterdam, The Netherlands}
\affiliation[b]{Department of Physics \& Astronomy, University of
  Manchester, Manchester M13 9PL, United Kingdom}
  \affiliation[c]{School of Physics and Astronomy, Monash University, Wellington Road, Clayton, VIC-3800, Australia}
\affiliation[d]{Rudolf Peierls Centre for Theoretical Physics, Clarendon
  	Laboratory, Parks Road, University of Oxford, Oxford OX1 3PU, UK}
\affiliation[e]{CERN, Theoretical Physics Department, CH-1211 Geneva
	23, Switzerland}

\emailAdd{mbeekvel@nikhef.nl}
\emailAdd{mrinal.dasgupta@manchester.ac.uk}
\emailAdd{basem.el-menoufi@monash.edu}
\emailAdd{jack.helliwell@physics.ox.ac.uk}
\emailAdd{alexander.karlberg@cern.ch}
\emailAdd{pier.monni@cern.ch}

\preprint{CERN-TH-2023-224, OUTP-24-02P, Nikhef 2024-001}

\abstract{We study the collinear fragmentation of highly energetic
  jets defined with a small jet radius.  In particular, we investigate
  how the corresponding fragmentation functions differ from their
  hadronic counterpart defined in the common $\overline{\rm MS}$
  scheme.
  We find that the anomalous dimensions governing the perturbative
  evolution of the two fragmentation functions differ starting at the
  two loop order. We compute for the first time the new anomalous
  dimensions at two loops and confirm our predictions by comparing the
  inclusive small-$R$ jet spectrum against a fixed order perturbative
  calculation at ${\cal O}(\alpha_s^2)$.
  To investigate the dependence of the anomalous dimension on the
  kinematic cutoff variable, we study the fragmentation functions of
  Cambridge jets defined with a transverse momentum cutoff as opposed
  to an angular cutoff $R$.
  We further study the evolution of the small-$R$ fragmentation
  function with an alternative cutoff scale, proportional to $z R$,
  representing the maximum possible transverse momentum of emissions
  within a jet.
  In these cases we find that the two-loop anomalous dimensions
  coincide with the $\overline{\rm MS}$ DGLAP ones, highlighting a
  correspondence between the $\overline{\rm MS}$ scheme and a
  transverse-momentum cutoff.}
  
\begin{document}

\setlength{\parskip}{0pt}
\maketitle
\flushbottom
%

\section{Introduction}
\label{sec:intro}

The increasing precision reached by the LHC experiments in the
measurement of the properties of QCD jets necessitates higher accuracy
in perturbative calculations of jet observables. In addition to
higher-order fixed-order calculations, multi-scale observables commonly
also require the resummation of logarithmically enhanced radiative
corrections of infrared origin.
In the latter context, we consider the problem of timelike collinear
fragmentation of hard partons produced in a high-energy scattering
process.
In this regime we are concerned with observables sensitive solely to
collinear logarithmic corrections, where the leading logarithms (LL)
are single logarithms $\alpha_s^n L^n$.  Observables of this class are
common in collider physics and have been extensively studied in the
literature (see
e.g.~\cite{Gribov:1972ri,Dokshitzer:1977sg,Altarelli:1977zs,Furmanski:1980cm,Curci:1980uw,Jain:2011xz,Alioli:2013hba,Chang:2013rca,Ritzmann:2014mka,Dasgupta:2014yra,Banfi:2015pju,Dasgupta:2016bnd,Kang:2016mcy,Chen:2020adz,Chen:2021gdk,Dasgupta:2021hbh,Dasgupta:2022fim,Chen:2022jhb,Chen:2022muj,Liu:2022wop,Chen:2023zlx,vanBeekveld:2023lsa,Cao:2023oef}).

For such single-logarithmic observables, a next-to-leading-logarithmic
(NLL) calculation would capture the next-to-single logarithmic tower
of corrections of order $\alpha_s^n L^{n-1}$.  Here the general
formulation of an algorithm to carry out NLL resummations requires the
consistent inclusion of $1\to 3$ splitting functions at tree-level
along with the one-loop corrections to the $1\to 2$ splitting
functions.  A possible theoretical approach to such resummation
algorithms can be formulated using the language of generating
functionals~\cite{Konishi:1979cb,Bassetto:1983mvz,Dokshitzer:1991wu},
tailored to this class of fragmentation problems in
Refs.~\cite{Dasgupta:2014yra,vanBeekveld:2023lsa}.
Such a formulation is rather powerful when trying to connect the field
of QCD resummation with that of parton showers, with which the
generating functional method is intimately linked.

An important aspect of the development of novel resummation techniques
is testing them across observables sensitive to different aspects of
fragmentation.
For example, Ref.~\cite{vanBeekveld:2023lsa} considered the NLL
calculation of a family of groomed angularities and fractional moments
of energy correlators. These observables are of a more inclusive kind
in that they are defined by integrating over the momentum fraction $z$
of a given collinear splitting.
As a next natural step, we wish to consider observables that are
\textit{differential} in $z$, to probe more deeply the
structure of our formulation.
Observables belonging to this class are the common NLO fragmentation
functions (FFs) measured either on final state
hadrons~\cite{Furmanski:1980cm,Curci:1980uw} or on final state jets
clustered with a small jet radius $R\ll 1$, which have been studied
for instance in
Refs.~\cite{Dasgupta:2014yra,Dasgupta:2016bnd,Kang:2016mcy}. These
observables are sensitive to both the $z$ dependence of the novel
anomalous dimension ${\cal B}_2^f(z)$ computed for quark and gluon
jets respectively in
Refs.~\cite{Dasgupta:2021hbh,vanBeekveld:2023lsa}, as well as to the
full triple-collinear $1\to 3$ squared splitting amplitudes
\cite{Campbell:1997hg,Catani:1998nv}.

In the process of testing our formulation on such observables we have
uncovered an important conceptual subtlety connected to FFs defined
with an angular cutoff $R$. In particular, we find that the anomalous
dimensions of small-$R$ FFs differ from the standard DGLAP ones
starting at the two loop order ${\cal O}(\alpha_s^2)$.
To the best of our knowledge there has been no previous calculation of
the two loop anomalous dimensions for small-$R$ jets, in previous
literature the answer at this order was conjectured from the form of
the evolution equation (see e.g. Ref.~\cite{Kang:2016mcy}).

Interestingly, the two-loop difference between these anomalous
dimensions takes a rather simple analytic form with a universal
structure across flavour channels.
Moreover, the term that breaks the correspondence between small-$R$
jets and timelike DGLAP evolution is of the same form and related
analytic origin as the term identified first by Curci, Furmanski, and
Petronzio~\cite{Curci:1980uw} responsible for breaking of the
Gribov-Lipatov reciprocity
relation~\cite{Gribov:1972rt,Mueller:1983js,Basso:2006nk,Dokshitzer:2006nm}
between timelike and spacelike anomalous dimension in the non-singlet
flavour channel.
In Ref.~\cite{Dokshitzer:2005bf} it was argued that terms of this type
are linked to the choice of the evolution cutoff (see also related
discussions in Ref.~\cite{Neill:2020bwv}). It is therefore also
interesting to explore the link with FFs defined by other types of
kinematical cuts, e.g.~a transverse momentum.
In order to investigate this observation, we compute the two-loop
evolution kernels for Cambridge jets defined with a transverse
momentum $y_{\rm cut}$ cutoff. We further compute a variant of the
small-$R$ FF, where the kinematic cut on the radiation is proportional
to $z\,R$ (with $z$ being the energy fraction of the jet), related to
the maximum possible transverse momentum of emissions within a
small-$R$ jet.
Remarkably, in both of these cases we find that the two-loop anomalous
dimensions now coincide with DGLAP. This result is critical to
understand how to reproduce DGLAP evolution beyond LO in parton
showers, which inevitably use a kinematical cutoff.\footnote{See also
  Ref.~\cite{Jadach:2011kc} and related work for investigations in a
  related direction.}

In this article we present the new anomalous dimensions for small-$R$
FFs at the two-loop order and confirm our predictions by comparing to
an exact numerical fixed order calculation at ${\cal O}(\alpha_s^2)$.
The paper is organised as follows. In Sec.~\ref{sec:FF} we study the
$\overline{\rm MS}$ fragmentation function and we analyse the effect
of placing an angular cutoff $R\ll 1$. We then present a simple recipe
to derive the small-$R$ anomalous dimensions at the two-loop order.
In Sec.~\ref{sec:tests} we consider the inclusive small-R jet spectrum
as a case study and test our findings against a fixed-order
calculation at ${\cal O}(\alpha_s^2)$ using the \textsc{Event2}
program~\cite{Catani:1996vz}. To investigate the dependence of the
anomalous dimensions on the kinematic cutoff, we also report the study
of the FF at ${\cal O}(\alpha_s^2)$ with a transverse momentum cutoff
in Sec.~\ref{sec:small-ycut}, as well as the study of the small-$R$
jet FF with a modified transverse-momentum-like cutoff scale.
In these cases the anomalous dimensions coincide with the DGLAP ones,
highlighting an important correspondence between the
$\overline{\rm MS}$ scheme and a transverse-momentum cutoff at the NLL
order.
Finally, our conclusions are presented in
Sec.~\ref{sec:conclusions}. We also include three appendices which
present: some relevant technical details in
App.~\ref{app:phase-space}, our full set of modifications to the DGLAP
anomalous dimensions in App.~\ref{app:ADs}, and a derivation of the
anomalous dimensions from the formalism of
Ref.~\cite{vanBeekveld:2023lsa} in App.~\ref{app:B2-derivation}.

\section{The inclusive microjet spectrum and small-$R$ fragmentation
  function}
\label{sec:FF}
We consider the inclusive small-$R$ jets (microjets) fragmentation
function which, for illustrative purposes, we define in the process
$e^+e^-\to$ jets. The cross section differential in the energy
fraction $z$ carried by a microjet is given, up to power corrections,
by
\begin{equation}\label{eq:small-R-FF}
\frac{1}{\sigma_0}\frac{d\sigma^{\rm jet}}{d z} \equiv \sum_{i={q,\bar{q},g}}
\int_{z}^1 \frac{d\xi}{\xi} C^{\rm jet}_i\left(\xi,\mu,Q\right) D^{\rm
  jet}_i\left(\frac{z}{\xi},\mu,E\,R\right)\,,
\end{equation}
with $\sigma_0$ denoting the Born cross section for
$e^+e^-\to q\bar{q}$. The matching coefficient
$C^{\rm jet}_i\left(\xi,\mu,Q\right)$ admits a fixed order
perturbative expansion in $\alpha_s(\mu)$. We
take $\mu$ to be of the order of the centre-of-mass energy $Q$
and more precisely we will set $\mu=E=Q/2$, i.e.~the energy of each
hemisphere. For this scale, the fragmentation function
$D^{\rm jet}_i\left(\frac{z}{\xi}, \mu,E\,R\right)$ resums the
logarithms of the small jet radius. Its evolution with the
factorisation scale $\mu$ is perturbative as long as
$ Q\,R \gg \Lambda_{\rm QCD}$ and it is governed by the equation
\sloppy
\begin{align}\label{eq:Djet}
\frac{d D^{\rm
  jet}_k\left(z, \mu,E\,R\right)}{d\ln\mu^2} =
  \sum_{i}\int_{z}^1\frac{d\xi}{\xi} \hat{P}_{ik}\left(\frac{z}{\xi},\mu,E\,R\right) D^{\rm
  jet}_i \left(\xi, \mu,E\,R\right)\,.
\end{align}
Here the index $i$ runs over all active quarks, antiquarks, and
gluons.
The main result of this article is that, up to NLL the anomalous
dimension $\hat{P}$ admits a perturbative expansion of the form
\begin{align}\label{eq:Pik}
\hat{P}_{ik}\left(z,\mu,E\,R\right) &= \frac{\alpha_s(\mu^2)}{2\pi}
  \left(\hat{P}_{ik}^{(0)}(z)+\frac{\alpha_s(\mu^2)}{2\pi}
    \hat{P}_{ik}^{(1),\,{\rm AP}}(z) -\frac{\alpha_s(E^2R^2)}{2\pi}
                                       \delta\hat{P}_{ik}^{(1)}+ {\cal O}({\rm NNLL})\right) \notag\\
  &=\frac{\alpha_s(\mu^2)}{2\pi}\left(
\hat{P}^{(0)}_{ik}\left(z\right) +
\frac{\alpha_s(\mu^2)}{2\pi}\hat{P}^{(1)}_{ik}\left(z\right) \right.\notag\\&\qquad\qquad\qquad\left.-\left(\frac{\alpha_s(E^2R^2)}{2\pi}-\frac{\alpha_s(\mu^2)}{2\pi}\right)
\delta\hat{P}^{(1)}_{ik}\left(z\right) + {\cal O}({\rm NNLL})\right)\,.
\end{align}
As we will show below, while at one loop the anomalous dimension
$\hat{P}^{(0)}_{ik}$ agrees with the Altarelli-Parisi one, which we
denote with $\hat{P}_{ik}^{(n),\,{\rm AP}}(z)$ here, this is not true
any longer at higher orders. In particular, we will show that
$\hat{P}_{ik}^{(1)}(z)=\hat{P}_{ik}^{(1),\,{\rm
    AP}}(z)-\delta\hat{P}_{ik}^{(1)}(z)$, where the new term
$\delta\hat{P}_{ik}^{(1)}(z)$ is a consequence of clustering
effects, which are absent in conventional $\MSbar$
evolution of the FFs.
Eq.~\eqref{eq:Djet} is solved as a path-ordered exponential describing
the evolution between a hard scale $\mu=E$ and a low scale $\mu=E\,R$,
at which the solution is convoluted with a boundary condition
$D^{\rm jet}_k\left(z, E\,R,E\,R\right)$. A LL resummation of the FF
would require the one-loop kernels, while the two-loop kernels as well
as the one-loop correction to the boundary condition are required at
NLL.

To define the small-$R$ jets we work with the \textit{inclusive}
generalised-$k_t$ family of algorithms~\cite{Cacciari:2011ma}, which
cluster sequentially proto-jets according to the distance measure
\begin{align}
d_{ij} &= 2 \,\min\left(\frac{E^2_i}{Q^2},
         \frac{E^2_j}{Q^2}\right)^p\,(1-\cos\theta_{ij})\,,\notag\\
  d_{iB} &= \frac{E^{2\,p}_i}{Q^{2\,p}}\,R^2\,,
\end{align}
where $E_i$ denotes the energy of proto-jet $i$ and $\theta_{ij}$ is
the angle between proto-jets $i$ and $j$. The algorithm proceeds by
finding the smallest between all $d_{ij}$ and $d_{iB}$ in the event
and recombining either $i$ and $j$ if $d_{ij}$ is the smallest
distance or promoting $i$ to a jet if $d_{iB}$ is the smallest
distance.
The proto-jets are recombined according to the $E$ scheme, in which
the four momenta are added together. The parameter $p$ identifies the
algorithm, with $p=1$ corresponding to the $k_t$
algorithm~\cite{Catani:1993hr}, $p=-1$ to
anti-$k_t$~\cite{Cacciari:2008gp} and $p=0$ to the Cambridge/Aachen
(C/A) algorithm~\cite{Dokshitzer:1997in,Wobisch:1998wt}. Though we
explicitly study the Cambridge-Aachen algorithm, at NLL the
considerations that follow will hold for all of the three
algorithms. This is because in the small-$R$ limit, differences
between these three algorithms only arise in kinematic configurations
that contribute starting at NNLL. In the case of
SISCone~\cite{Salam:2007xv}, the only difference at the NLL order is
entirely encoded in the boundary condition to the evolution
equation~\eqref{eq:Djet}, while the anomalous dimensions are identical
to those of the inclusive generalised-$k_t$ family of jet algorithms.

The all-order structure of Eq.~\eqref{eq:Djet} is closely related to
that of the standard fragmentation function, governed by the DGLAP
equation~\cite{Gribov:1972ri,Dokshitzer:1977sg,Altarelli:1977zs}.
However, because of a subtlety we will discuss in the following, the
anomalous dimensions governing the evolution of $D^{\rm jet}_i$ will
differ from the DGLAP anomalous dimension describing the evolution of
the $\overline{\rm MS}$ fragmentation function starting at the
two-loop order. This implies that while
$\hat{P}^{(0)}_{ik}\left(z\right)=\hat{P}_{ik}^{(0),\,{\rm AP}}(z)$ in
Eq.~\eqref{eq:Pik}, this is not true at higher orders in $\alpha_s$.
In the following section we will derive the one-loop matching
coefficients needed at NLL, while in Sec.~\ref{sec:two-loop} we will
compute the new two-loop anomalous dimension. Finally, in
Sec.~\ref{sec:running} we will discuss the running coupling of the new
term $\delta\hat{P}_{ik}^{(1)}(z)$ along with the physical
interpretation of this effect. Appendix~\ref{app:B2-derivation}
reports a derivation of the final evolution equation~\eqref{eq:Djet}
using the formalism of Ref.~\cite{vanBeekveld:2023lsa}.

\subsection{One loop matching coefficients and boundary conditions}
We start with a simple ${\cal O}(\alpha_s)$ calculation of
Eq.~\eqref{eq:small-R-FF}, which will allow us to extract the one-loop
matching coefficients $C_i^{\rm jet}$ as well as the boundary
condition to Eq.~\eqref{eq:Djet}, both of which are needed for a NLL
calculation.

At ${\cal O}(\alpha_s)$, we start with the production of a $q\bar{q}g$
final state, and consider the action of a clustering algorithm into
small-$R$ jets. Since we work in the limit $R\ll 1$ we only need to
consider kinematic configurations enhanced by a collinear singularity
(hence producing logarithms of $R$). This simply amounts to
configurations in which the gluon is collinear to either of the two
quarks.
The resulting calculation is thus identical to that of the standard
fragmentation function in the regions $\theta_{q g}^2 > R^2$ and
$\theta_{\bar{q} g}^2 > R^2$, but it gets modified when the gluon is
recombined with one of the quarks. At the one-loop order, this
modification effectively replaces the ${\cal O}(\epsilon^{-1})$ pole
of the $\overline{\rm MS}$ fragmentation function with $\ln 1/R$ and
it will produce a low-scale non-logarithmic term, which serves as a
boundary condition at $\mu=E\,R$. We obtain
\begin{align}
D^{\rm jet}_i\left(z,\mu, E\,R\right) = \delta(1-z) + \frac{\alpha_s(\mu)}{2\pi}\,D^{{\rm
  jet}\,(0)}_i\left(z,\mu, E\,R\right) + {\cal O}(\alpha_s^2)\,,
\end{align}
where
\begin{align}\label{eq:bc-D}
D^{{\rm  jet}\,(0)}_q\left(z,E\,R, E\,R\right)
  &=-2\,C_F\,(1+z^2)\,\left(\frac{\ln(1-z)}{1-z}\right)_+\notag\\
  & + C_F\,\left(-1+2\,
    \left(2-\frac{2}{z}-z\right)\,\ln\left(1-z\right)+\left(6-\frac{4}{(1-z)\,z}\right)\,\ln
    z\right)\notag\\
  &+C_F\,\left(\frac{13}{2}-\frac{2}{3}\,\pi^2\right)\,\delta(1-z)\,,\\
D^{{\rm  jet}\,(0)}_{\bar q}\left(z, E\,R, E\,R\right) &= D^{{\rm
                                                        jet}\,(0)}_q\left(z, E\,R,
                                                       E\, R\right) \,,
\end{align}
\begin{align}
D^{{\rm  jet}\,(0)}_g\left(z, E\,R, E\,R\right)
  &=2\,\left(\hat{P}^{(0)}_{gg}(z)+2\,n_f\,\hat{P}^{(0)}_{qg}(z)\right)\ln\frac{1}{z}-4\,C_A\,\frac{\left(1-z+z^2\right)^2}{z}\, \left(\frac{\ln(1-z)}{1-z}\right)_+\notag\\
  & - 4\, n_f\,\hat{P}^{(0)}_{qg}(z)\,\ln(1-z)-4\,T_R\,n_f\,z\,(1-z)\notag\\
  &+\left(C_A\,\left(\frac{67}{9}-\frac{2}{3}\,\pi^2\right)-T_R\,n_f\,\frac{23}{9}\right)\,\delta(1-z)\,.
\end{align}

Here we defined the regularised splitting kernels
$\hat{P}^{(0)}_{gg}(z)$ and $\hat{P}^{(0)}_{qg}(z)$ as
\begin{equation}
\hat{P}^{(0)}_{qg} = T_R \left(z^2+(1-z)^2\right)\,,\quad \hat{P}^{(0)}_{gg} =
2\,C_A\,\left(\frac{z}{(1-z)_+} + \frac{1-z}{z} + z\,(1-z)\right) + b_0\,\delta(1-z)\,,
\end{equation}
with $b_0= 11/6 \,C_A - 2/3\,T_R\,n_f$, $C_F = 4/3$, $T_R = 1/2$
and $C_A=3$, while $n_f$ is the number of light-quark flavours,
assumed to be $5$ here.

The above expressions agree with the results of
Ref.~\cite{Kang:2016mcy}.\footnote{We find a different result for the
  SISCone algorithm already at ${\cal O}(\alpha_s)$, which can be
  traced back to an initial error in
  Refs.~\cite{Mukherjee:2012uz,Kang:2016mcy}, later corrected in
  Ref.~\cite{Kang:2017mda}. Our results agree with the latter, and
  therefore we refrain from reporting them here.}
Finally, the matching coefficients $C_i^{\rm jet}$ are simply obtained
by matching the calculation of $D^{\rm jet}_i$ to the full QCD
calculation at one loop \cite{Curci:1980uw}, the result is also given
in eq. (2.16) of \cite{Nason:1993xx}.
At this order, they agree with the sum of longitudinal and transverse
coefficient functions entering the $\overline{\rm MS}$ fragmentation
function which were first computed in Ref.~\cite{Curci:1980uw}. These
read
\begin{align}
C^{\rm jet}_i\left(z,\mu,Q\right) = \delta(1-z) + \frac{\alpha_s(\mu)}{2\pi}\,C^{{\rm
  jet}\,(0)}_i\left(z, \mu,Q\right) + {\cal O}(\alpha_s^2)\,,
\end{align}
where, setting $\mu=E$,
\begin{align}
C^{{\rm
  jet}\,(0)}_q(z,E,Q) &= C_F\,(1+z^2)\, \left(\frac{\ln(1-z)}{1-z}\right)_+\\
&- C_F\,\left(\frac{3}{2}-2 \,(1+z^2)\ln 2\right)
                       \left(\frac{1}{1-z}\right)_+\notag\\
                        & +\frac{C_F}{2}\,\left(5-3\,z+4\,\frac{(1+z^2)}{1-z}\, \ln
                          z\right)-C_F\,\left(\frac{9}{2}-\frac{2}{3}\,\pi^2-3\ln2\right)\delta(1-z)\,,\notag
\end{align}
\begin{align}
C^{{\rm
  jet}\,(0)}_{\bar q}(z, E,Q) &=C^{{\rm
  jet}\,(0)}_q(z, E,Q)\,,\\
C^{{\rm
  jet}\,(0)}_g(z, E,Q)
                     &=2\,C_F\,\frac{2-2\,z+z^2}{z}\,\left(\ln(1-z)+2\,\ln z+2\,\ln2\right)\,.
\end{align}

\subsection{The two-loop anomalous dimensions}
\label{sec:two-loop}
We can now proceed with the analysis at the two-loop order, by
calculating the splitting kernels $\hat{P}^{(1)}_{ik}$. For ease of
presentation, we use a simple method for the calculation of the
two-loop anomalous dimension in the non-singlet (NS) flavour channel,
where we target the $C_F^2$ colour structure and hence set for the
time being $C_A=n_f=0$.
It will then be evident how this argument can be generalised to the
remaining colour and flavour channels, allowing us to derive the full
anomalous dimension at two loops.

\subsubsection{The $C_F^2$ contribution to the NS channel}
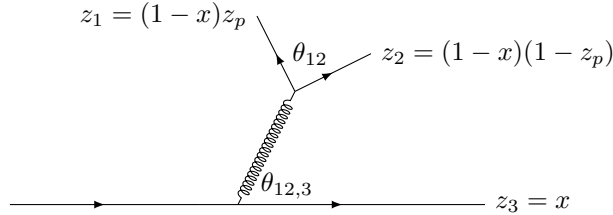
\begin{figure}[h]
	\centering
	\begin{tikzpicture}[scale=2.5] 
		
		\coordinate (bq1) at (0,0);
		\coordinate (bq2) at (1,0); 
		\coordinate (bq3) at (2.5,0);
		
		\coordinate (bg1) at (1.2,0);
		\coordinate (eg1) at (1.5,0.6);
		\coordinate (bg2) at (1.5,0.6);
		\coordinate (eg2) at (1.3,1.0);
		\coordinate (bg3) at (1.5,0.6);
		\coordinate (eg3) at (1.9,0.8);
		
		\draw [quark] (bq1) -- (bq2);
		\draw [gluon] (bg1) -- (eg1) ;
		\node at (1.4,0.1) {\small \,\,\,\,\,$\theta_{12,3}$};
		\draw [quark] (bg2) -- (eg2) node [pos=1,left] {\small $z_1=(1-x)z_p$};
		\node at (1.58,0.8) {\small $\theta_{12}$};
		\draw [quark] (bg3) -- (eg3) node [pos=1,right] {\small $z_2=(1-x)(1-z_p)$};
		\draw [quark] (bq2) -- (bq3) node
		[pos=1,right] {\small $z_3 = x$} ;
		
	\end{tikzpicture}
	\caption{Phase space parametrisation for the production of two
          identical quarks, labelled by particles 1 and 3.}
	\label{fig:a-splits}
\end{figure}

\begin{figure}[h]
	\centering
	\begin{tikzpicture}[scale=2.5] 
		
		\coordinate (bq1) at (0,0);
		\coordinate (bq2) at (1,0); 
		\coordinate (bq3) at (2.5,0);

		\coordinate (bg1) at (.8,0);
		\coordinate (eg1) at (1.2,0.6);
		\coordinate (bg2) at (1.65,0);
		\coordinate (eg2) at (2.05,0.6);
		
		\draw [quark] (bq1) -- (bq2);
		\draw [gluon] (bg1) -- (eg1) node [pos=1,right] {\small $z_1=1-x$} ;
		\node at (1.02,0.1) {\small \,\,\,\,\,$\theta_{1,23}$};
		\draw [gluon] (bg2) -- (eg2) node [pos=1,right] {\small $z_2=x(1-z_p)$};
		\node at (1.95,0.1) {\small $\theta_{23}$};
		\draw [quark] (bq2) -- (bq3) node
		[pos=1,right] {\small $z_3 = x z_p$};

	\end{tikzpicture}
        \caption{Phase space parametrisation for the radiation of two
          independent gluons off a quark line.}\label{fig:b-splits}
\end{figure}
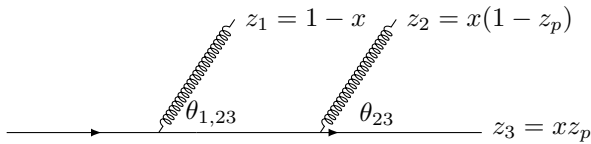

Since for the time being we are interested in the $C_F^2$ contribution
to the NS channel, we must take into account the abelian collinear
splittings depicted in Fig.~\ref{fig:a-splits} ($q\to q q\bar{q}$ --
with identical quark flavours) and Fig.~\ref{fig:b-splits}
($q\to q gg$), whose spin-averaged kernels $\langle P \rangle$ in the
triple collinear limit can be found in
Refs.~\cite{Campbell:1997hg,Catani:1998nv,Sborlini:2013jba,Braun-White:2022rtg}.
To extract the $C_F^2$ contribution to the two loop splitting kernel
$\hat{P}_{qq}^{(1)}$, it is convenient to start from the
identical-quark contribution to the standard FF (i.e.~not for
small-$R$ jets) depicted in Fig.~\ref{fig:a-splits}. For $z\neq 1$
this reads
\begin{align}\label{eq:I-0}
I^{\rm id.}_{C_F^2}(\epsilon, z) &\equiv \,\frac{1}{2!}\int
  d\Phi_{3}^{(A)} \frac{(8 \pi\, \bar{\alpha}_s\mu^{2\epsilon})^2}{s_{123}^2}\,\langle
    P\rangle_{C_F\,(C_F-C_A/2)}\left(\delta(z-(1-x) z_p) +
    \delta(x-z)\right)\,,
\end{align}
where the identical-fermion contribution to the $1\to 3$ splitting
function $\langle P\rangle_{C_F\,(C_F-C_A/2)} $ is given in Eq.~(31) of
Ref.~\cite{Catani:1998nv}.
The coupling $\alpha_s$ is renormalised in the $\overline{\rm MS}$
scheme and we used the short-hand notation
$\bar{\alpha}_s \equiv S^{-1}_\epsilon\,\alpha_s$, with
$S_\epsilon = (4\pi)^\epsilon \, e^{-\epsilon\gamma_E}$.
The phase space measure $d\Phi_{3}^{(A)}$ is defined in
App.~\ref{app:phase-space}, $s_{123}$ is the squared invariant mass of
the collinear three-parton system, and the two $\delta$ functions
correspond to fixing the energy fraction of either of the two final
state quarks contributing to the NS FF. The two contributions are
identical and exactly cancel the $1/2!$ symmetry factor. The
integration is straightforward and the result can be found, for
instance, in Ref.~\cite{Dasgupta:2021hbh} (and specifically by
multiplying Eq.~(3.7) in this reference by a factor $-1/\epsilon$).
The single pole directly contributes to the NLO DGLAP anomalous
dimension in the non-singlet channel.

The next step is then to consider the same contribution to the
small-$R$ FF. An important property of the integral~\eqref{eq:I-0} is
that it contributes only a single pole to the FF, and it does not
receive a contribution from virtual corrections. This is reflected in
the fact that there is no collinear singularity when the final state
particles are strongly ordered in angle. We then consider applying the
C/A clustering which, in the limit $R\ll 1$, will only modify
Eq.~\eqref{eq:I-0} in configurations where the relative angles become
small. Due to the absence of a collinear enhancement in such
configurations, the contribution of such regions is power suppressed
in $R^2$. This immediately implies that the contribution from this
channel to the small-$R$ two loop anomalous dimension
$\hat{P}^{(1)}_{qq}$ will coincide with the corresponding DGLAP
counterpart.

Next, we consider the abelian channel depicted in
Fig.~\ref{fig:b-splits}. As before we start by examining its
contribution to the hadronic FF. One can single out the two loop
correction to the Altarelli-Parisi splitting kernel by calculating the
following combination
\begin{align}\label{eq:I-1}
  I^{\rm ab.}_{C_F^2}(z,\epsilon) &\equiv \,\frac{1}{2!}\int
                                     d\Phi_{3}^{(B)} \frac{(8 \pi\, \bar{\alpha}_s\mu^{2\epsilon})^2}{s_{123}^2}\,\langle
                                     P\rangle_{C_F^2} 
                                     \delta(z - x z_p) \notag\\&+\int \frac{d\theta^2}{\theta^{2}}\,d x\, \frac{\alpha_s^2}{(2\pi)^2}\,{\cal
                                     V}^{(1),\,C_F^2}_{q\to q
                                     g}(x, \theta, \epsilon)\,\delta(x-z)\notag\\
                                   &-\frac{1}{2!} \int d\Phi_{2}^2\, \frac{(8 \pi\,\bar{\alpha}_s\mu^{2\epsilon})^2}{E^4}\frac{P^{(0)}_{qq}(x,\epsilon)}{\theta_{13}^2}
                                     \frac{P^{(0)}_{qq}(z_p,\epsilon)}{\theta_{23}^2} \delta(z-x z_p) \,,\qquad
                                     z\neq 1\,,
\end{align}
where the phase space measure $d\Phi_{3}^{(B)}$ is given in
App.~\ref{app:phase-space}, and the abelian contribution to the
$1\to 3$ splitting function $\langle P\rangle_{C_F^2}$ can be found in
Eq.~(33) of Ref.~\cite{Catani:1998nv}.\footnote{We note that the
  overall $C_F^2$ colour factor is included in
  $\langle P\rangle_{C_F^2}$ in our notation.} 
Finally, the one-loop correction to the $1\to 2$ splitting function
${\cal V}^{(1),\,C_F^2}_{q\to q g}$ was calculated in
Ref.~\cite{Sborlini:2013jba} and in the convention of
Eq.~\eqref{eq:I-1} is given in the r.h.s.\ of Eq.~(3.42) of
Ref.~\cite{Dasgupta:2021hbh} with a factor of $\alpha_s^2/(2\pi)^2$
removed.
The term in the third line of Eq.~\eqref{eq:I-1} has the role of
subtracting the LL terms from the first two lines, hence cancelling
all contributions originating from configurations strongly ordered in
angle. The $d\Phi_{2}^2$ phase space measure is defined starting from
the iteration of the $1\to 2$ phase space and it is given by
\begin{equation}
	\label{eq:dPhi2}
d\Phi_{2}^2\equiv \frac{E^{4-4\epsilon}}{(4\pi)^{4-2\epsilon}\Gamma(1-\epsilon)^2}\, d\theta_{13}^2\, d\theta_{23}^2\, dx\, d z_p\,\theta_{13}^{-2\epsilon}\theta_{23}^{-2\epsilon} (x(1-x))^{-2\epsilon} (z_p(1-z_p))^{-2\epsilon}\,.
\end{equation}
We can recast Eq.~\eqref{eq:I-1} as (for $z\neq 1$)
\begin{align}\label{eq:I-2}
I^{\rm ab.}_{C_F^2}&( z, \epsilon) = \,\frac{1}{2!}\int
  d\Phi_{3}^{(B)}(8 \pi \bar{\alpha}_s\mu^{2\epsilon})^2 \left(\frac{1}{s_{123}^2}\,\langle
                        P\rangle_{C_F^2} - \frac{{\cal J}(x,z_p)}{E^4}\,\frac{P^{(0)}_{qq}(x,\epsilon)}{\theta_{13}^2}
    \frac{P^{(0)}_{qq}(z_p,\epsilon)}{\theta_{23}^2} \right)
\delta(z - x z_p) \notag\\&+\int \frac{d\theta^2}{\theta^{2}}\,d x\, \frac{\alpha_s^2}{(2\pi)^2}\,{\cal
                                     V}^{(1),\,C_F^2}_{q\to q
                                     g}(x, \theta, \epsilon)\,\delta(x-z)\notag\\
    &+\frac{1}{2!} \left(\int
  d\Phi_{3}^{(B)} \,{\cal J}(x,z_p)\, - \int d\Phi_{2}^2\right)\, \frac{(8 \pi\,\bar{\alpha}_s\mu^{2\epsilon})^2}{E^4}\, \frac{P^{(0)}_{qq}(x,\epsilon)}{\theta_{13}^2}
    \frac{P^{(0)}_{qq}(z_p,\epsilon)}{\theta_{23}^2}\,\delta(z-x z_p) \,,
\end{align}
where the prefactor ${\cal J}(x,z_p)$ is given by
\begin{equation}\label{eq:JPS}
{\cal J}(x,z_p) \equiv  \frac{1}{z_1 z_2 z_3 x} = \frac{1}{(1-x) x^3 (1-z_p) z_p }\,.
\end{equation}

To make contact with the small-$R$ FF, we partition the
three-body phase space according to the Cambridge-Aachen algorithm,
that is we trade the $1/2!$ in the double-real radiation integrals by
an angular-ordering condition $\Theta(\theta_{13}^2-\theta_{23}^2)$.
In considering the action of the jet algorithm we are allowed to
neglect two effects which are subleading for the present analysis. The
first is the contribution from configurations where the jet algorithm
clusters emissions in regions of phase space that are free of
collinear singularities (e.g. when the two gluons in
Fig.~\ref{fig:b-splits} are first clustered together). These
clustering corrections are power suppressed in $R^2$ and therefore can
be neglected in the small-$R$ limit.
The second effect that we can neglect in the calculation performed
below is the recoil of the jet axis due to recombination kinematics
following a clustering.
We have checked by explicit computation that this effect amounts to a
contribution to the boundary condition at ${\cal O}(\alpha_s^2)$, and
therefore is subleading w.r.t.~the accuracy considered here.

We now discuss the effect of the clustering on each of the terms in
Eq.~\eqref{eq:I-2}.
Due to the subtraction of the strongly-ordered regime, the first term
in Eq.~\eqref{eq:I-2} enjoys the same properties as the
identical-quark correction in Eq.~\eqref{eq:I-0}, in that it is free
of soft and collinear divergences. For this reason, any region of
phase space where jet clustering is active gives only a power
suppressed contribution ${\cal O}(R^2)$. This again implies that the
corresponding contribution to the splitting kernel
$\hat{P}^{(1)}_{qq}$ is the same for the DGLAP and small-$R$ cases.

We next consider the second and third line of Eq.~\eqref{eq:I-2}.
It is convenient to split the real phase space integrals by
introducing the partition of unity
\begin{equation}
1 = \Theta(\theta^2_{13}-R^2)+\Theta(R^2-\theta^2_{13})\,.
\end{equation}
Analogously, for the virtual corrections
${\cal V}^{(1),\,C_F^2}_{q\to q g}$ we insert
$1 = \Theta(\theta^2-R^2)+\Theta(R^2-\theta^2)$.  Accordingly, we can
write
\begin{equation}\label{eq:partition}
I^{\rm ab.}_{C_F^2}(z,\epsilon) = I^{\rm ab.}_{>R^2}(z,\epsilon) +
I^{\rm ab.}_{<R^2}(z,\epsilon)\,.
\end{equation}

We now consider the calculation of each of the above two terms,
starting with the contribution $I^{\rm ab.}_{<R^2}(z,\epsilon)$. In
this case, the sum of the second and third line of Eq.~\eqref{eq:I-2}
simply contributes a $\delta(1-z)$ to the small-$R$ FF, which does not
have any $\ln 1/R^2$ enhancement and hence amounts to a subleading,
NNLL boundary condition.
Secondly, we focus on the term $I^{\rm ab.}_{>R^2}(z,\epsilon)$. The
only source of difference between the small-$R$ and DGLAP anomalous
dimensions is due to the third line in Eq.~\eqref{eq:I-2} in a region
where $\theta_{23}^2 < R^2$ while $\theta_{13}^2 > R^2$. In this case,
we can directly calculate the difference between the DGLAP and
small-$R$ anomalous dimensions by evaluating the integral 
\begin{align}\label{eq:dP1+-}
  & \left(\int
  d\Phi_{3}^{(B)} \,{\cal J}(x,z_p)\, - \int d\Phi_{2}^2\right)\, \frac{(8 \pi\,\bar{\alpha}_s\mu^{2\epsilon})^2}{E^4}\, \frac{P^{(0)}_{qq}(x,\epsilon)}{\theta_{13}^2}
    \frac{P^{(0)}_{qq}(z_p,\epsilon)}{\theta_{23}^2}\, \Theta(\theta^2_{13}-R^2) \Theta(R^2-\theta^2_{23}) \notag\\&\times(\delta(z-x
  z_p)-\delta(z-x)) =
  \frac{\alpha_s^2}{(2\pi)^2}\left(\delta
  \hat{P}_{qq}^{(1)} (z)\ln\frac{1}{R^2}+{\cal O}(\epsilon)\right) \,,
\end{align}
where the two $\delta$ functions encode the difference between the
hadronic and small-$R$ jet FF.
In obtaining the above equation we have made use of the fact that the
contribution of the virtual corrections
${\cal V}^{(1),\,C_F^2}_{q\to q g}$ is common to both FFs and thus it
cancels.
Eq.~\eqref{eq:dP1+-} reveals that the non-singlet contribution to the
two-loop anomalous dimension for the small-$R$ FF can be obtained from
the corresponding DGLAP splitting kernel by subtracting the quantity
$\delta\hat{P}_{qq}^{(1)} (z)$, namely
\begin{align}\label{eq:Ptilde1}
  \hat{P}_{qq}^{(1)} (z) = \hat{P}_{qq}^{(1),\,{\rm AP}} (z) - \delta \hat{P}_{qq}^{(1)} (z) \,,
\end{align}
where we denote by $\hat{P}_{qq}^{(1),\,{\rm AP}} (z)$ the standard
DGLAP evolution kernels, and~\footnote{The standard convolution
  operator is defined as
  $f(z)\otimes g(z) \equiv \int_{z}^1d\,x/x\,f(x)\,g(z/x)$.}
\begin{align}\label{eq:result}
\delta \hat{P}_{qq}^{(1)} (z) &\equiv \left(2\,\ln z\, \hat{P}_{qq}^{(0)}\right)\otimes \hat{ P}_{qq}^{(0)} \notag\\
  & = -C_F^2\,\ln z \left(\frac{3\,z^2+1}{1-z} \ln z - 4
    \frac{1+z^2}{1-z}\ln(1-z) - \frac{z (4+z)+1}{1-z}\right)\,,
\end{align}
with
\begin{align}
\hat{P}^{(0)}_{qq}(z) &= C_F\,(1+z^2)\,\left(\frac{1}{1-z}\right)_+ +
                        C_F\,\frac32\,\delta(1-z)\,.
\end{align}
The remarkably simple structure of the result can be understood by
inspecting Eq.~\eqref{eq:dP1+-}. Here, the difference between the two
phase space measures $d\Phi_{3}^{(B)}$ and $d\Phi_{2}^2$ is of
${\cal O}(\epsilon\,\ln x)$, arising from the extra $x^{-2\epsilon}$
factor in the D-dimensional three-body phase space
$d\Phi_{3}^{(B)}$. This multiplies a $1/\epsilon$ pole of collinear
origin arising from the $\theta_{23}\to 0$ limit, giving a finite
leftover. Finally, the difference between the two $\delta$ functions
in Eq.~\eqref{eq:dP1+-} is reflected in the regularised splitting
functions in Eq.~\eqref{eq:result}.

The above result is the central observation of this article. It
highlights a difference between the anomalous dimension governing
standard DGLAP evolution and that of a fragmentation function defined
with an angular cutoff.
It is noteworthy that this term looks precisely like the difference
between the timelike and the spacelike two-loop splitting
kernels~\cite{Curci:1980uw} that breaks the reciprocity
relation~\cite{Mueller:1983js,Dokshitzer:2005bf,Basso:2006nk,Dokshitzer:2006nm}
between timelike and spacelike anomalous dimensions.
The origin of this term was previously found to be related to the
specific kinematic infrared cutoff (e.g.~angular vs.~transverse
momentum) in the collinear radiation~\cite{Dokshitzer:2006nm}. To
investigate this correspondence, in Sec.~\ref{sec:small-ycut} we will
calculate and test the two-loop anomalous dimensions for the
fragmentation functions of Cambridge jets, defined with a
transverse-momentum cutoff.

\subsubsection{Generalisation to all colour and flavour channels}
The arguments outlined above can be generalised to other
flavour channels. In general, starting at two loops, a given entry to
the anomalous dimension matrix for the small-$R$ FF will take the form
\begin{equation}\label{eq:tildeP}
\hat{P}^{(1)}_{ i k} = \hat{P}^{(1),\,{\rm AP}}_{ i k} - \delta\hat{P}^{(1)}_{ i k}\,.
\end{equation}
To calculate $\delta\hat{ P}^{(1)}_{ i k}$, we observe that it
originates from a sequence of two collinear splittings in which an
${\cal O}(\epsilon)$ contribution to the first splitting (which
amounts to the leading-order splitting function multiplied by a
$\ln z$ factor coming from the expansion of the $D$ dimensional phase
space measure) is convoluted with the ${\cal O}(\epsilon^{-1})$ pole
term of the second splitting (amounting to the corresponding
leading-order splitting function).
Concretely, we consider the sequence of $1\to 2$ splittings
originating from the fragmentation of a parton $A$, i.e.
\begin{equation}
  A \to B C \to (DE)\, C\,,\quad A \to B C \to B\, (FG)\,.
\end{equation}
The above sequences will then contribute to the following anomalous
dimensions
\begin{align}\label{eq:newADs}
  \delta\hat{ P}^{(1)}_{ EA} (z) \equiv \left(2\,\ln z\, \hat{P}_{BA}^{(0)}\right)\otimes 
  \hat{ P}_{EB}^{(0)} \,;&\quad \delta\hat{ P}^{(1)}_{ DA} (z) \equiv \left(2\,\ln z\,  \hat{P}_{BA}^{(0)}\right)\otimes
  \hat{ P}_{DB}^{(0)}\,,\notag\\
  \delta\hat{ P}^{(1)}_{ GA} (z) \equiv \left(2\,\ln z\,  \hat{P}_{CA}^{(0)}\right)\otimes
  \hat{ P}_{GC}^{(0)} \,;&\quad \delta\hat{ P}^{(1)}_{ FA} (z) \equiv \left(2\,\ln z\, \hat{P}_{CA}^{(0)}\right)\otimes
  \hat{ P}_{FC}^{(0)}\,,
\end{align}
where the sum over the intermediate states, i.e. over the indices $B$
and $C$, in each channel is understood.
The relevant corrections $\delta\hat{ P}^{(1)}_{ i k}$ at the two loop
order are reported in App.~\ref{app:ADs}.
A crucial property of the extra terms in the small-$R$ anomalous
dimensions is that they must satisfy the sum rules. Using the
expressions given in App.~\ref{app:ADs} we consistently find
\begin{align}
\int_0^1d\,z\,z\,\left(\delta\hat{ P}^{(1)}_{qq}(z) +\delta\hat{
  P}^{(1)}_{gq}(z) +\delta\hat{ P}^{(1)}_{\bar{q}q}(z)\right) &=
                0\,,\notag\\
 \int_0^1d\,z\,z\,\left(\delta\hat{ P}^{(1)}_{gg}(z) +\delta\hat{
  P}^{(1)}_{qg}(z) +\delta\hat{ P}^{(1)}_{\bar{q}g}(z)\right) &=
  0\,,\notag\\
 \int_0^1d\,z\,\left(\delta\hat{ P}^{(1)}_{qq}(z) -\delta\hat{ P}^{(1)}_{\bar{q}q}(z)\right) &=
                0\,.
\end{align}

\subsection{Running coupling effects beyond two loops}
\label{sec:running}
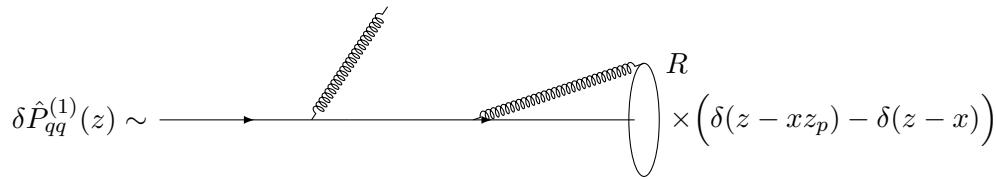
\begin{figure}[h]
	\centering
	\begin{tikzpicture}[scale=2.5] 
		
		\node at (-0.42, 0.018) {$\delta \hat{P}^{(1)}_{qq}(z) \sim$};
		\coordinate (bq1) at (0,0);
		\coordinate (bq2) at (1,0); 
		\coordinate (bq3) at (2.5,0);

		\coordinate (bg1) at (.8,0);
		\coordinate (eg1) at (1.2,0.6);
		\coordinate (bg2) at (1.65,0);
		\coordinate (eg2) at (2.55,0.3);
		
		\draw [quark] (bq1) -- (bq2);
		\draw [gluon] (bg1) -- (eg1) node [pos=1,right]{}; 
		\draw [gluon] (bg2) -- (eg2) node [pos=1,right]{}; 
		\node at (1.95,0.1) {}; 
		\node[label={[label distance=0.01cm]0:$R$}] at (2.55,0.3) {};
		\draw [quark] (bq2) -- (bq3) node
		[pos=1,right] {}; 
		\draw[] (2.55,0.0) ellipse (0.08cm and 0.3cm);
		\node at (3.55, 0.0) {$\times \Big(\delta(z - x z_p) - \delta(z - x)\Big)$};
	\end{tikzpicture}
	
	\caption{Clustering configuration giving rise to
          $\delta\hat{P}_{qq}^{(1)}$.}\label{fig:dPqq}
\end{figure}

In this section we will discuss the scale of the coupling multiplying
$\hat{P}^{(1)}_{ik}(z)$ in the evolution equation of the small-$R$ FFs
shown in Eq.~\eqref{eq:Djet}. Here we will present a simple physical
argument to justify the expression given in Eq.~\eqref{eq:Pik}. A full
derivation of this equation leading to the explicit scale of the
coupling for the term proportional to $\delta \hat{P}^{(1)}_{ik}(z)$
is reported in Appendix~\ref{app:B2-derivation}.
As shown explicitly in the previous section, the two loop anomalous
dimension $\hat{P}^{(1)}_{ik}(z)$ can be decomposed into the
difference between the DGLAP kernel
$\hat{P}^{(1),\,{\rm AP}}_{ik}(z)$ and $\delta\hat{P}^{(1)}_{ik}(z)$
given in Eq.~\eqref{eq:newADs}. The latter correction originates from
the change in the longitudinal momentum fraction of the final state
jet in a configuration in which one of the radiated partons clusters
with the jet. This scenario is depicted in Fig.~\ref{fig:dPqq} for the
$\delta \hat{P}_{qq}^{(1)}$ case.
In this configuration, the emission outside the jet will have a
coupling evaluated at a scale $\mu^2$ that runs between $E^2R^2$ and
$E^2$. Conversely, the scale of the coupling associated with the emission inside
the jet is bounded by $E^2R^2$ due to the constraint imposed by the jet
radius on the angle of the emission.
Therefore, the two terms defining $\hat{P}^{(1)}_{ik}(z)$ in
Eq.~\eqref{eq:tildeP} enter the NLL evolution equation evaluated at
two different scales, that is as
\begin{equation}
\alpha_s(\mu^2)\left(\alpha_s(\mu^2)\hat{P}^{(1),\,{\rm AP}}_{ i k} -\alpha_s(E^2R^2) \delta\hat{P}^{(1)}_{ i k}\right)\,.
\end{equation}
The above result is explicitly derived in Appendix~\ref{app:B2-derivation} using
the formalism of Ref.~\cite{vanBeekveld:2023lsa} and it justifies the
anomalous dimension given in Eq.~\eqref{eq:Pik}, where the term in the
second line has the role of changing the scale of the $
\delta\hat{P}^{(1)}_{ i k}$ term to $E^2R^2$.

\section{Fixed order test through ${\cal O}(\alpha_s^2)$}
\label{sec:tests}
To test our prediction for the inclusive micro-jet spectrum,
we compare the perturbative expansion of Eq.~\eqref{eq:small-R-FF} to
a fixed-order prediction obtained with the program
\textsc{Event2}~\cite{Catani:1996vz}.
We define the quantity
\begin{equation}
\Delta_i(z,R) \equiv
\frac{1}{\sigma_0}\,\left(\left.\frac{d\sigma^{\rm jet}}{d
      z}\right|^{(i)}_{\textsc{Event2}}-\left.\frac{d\sigma^{\rm jet}}{d
      z}\right|^{(i)}_{\rm Eq.~\eqref{eq:small-R-FF}}\right)\,,
\end{equation}
where the super-script $(i)$ indicates the perturbative order ${\cal
  O}(\alpha_s^i)$ of the expansion.

\begin{figure}[htbp]
	\centering
	\includegraphics[width=0.48\textwidth]{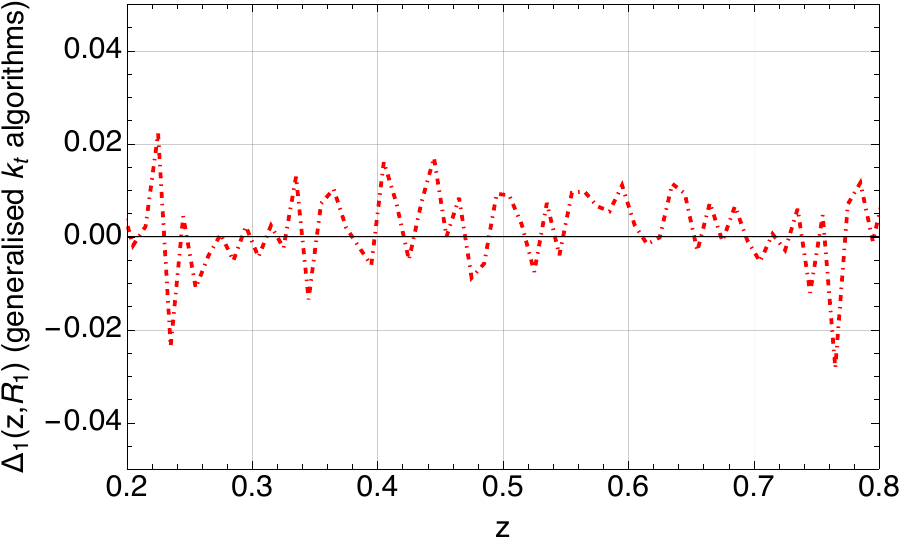}
      	\includegraphics[width=0.48\textwidth]{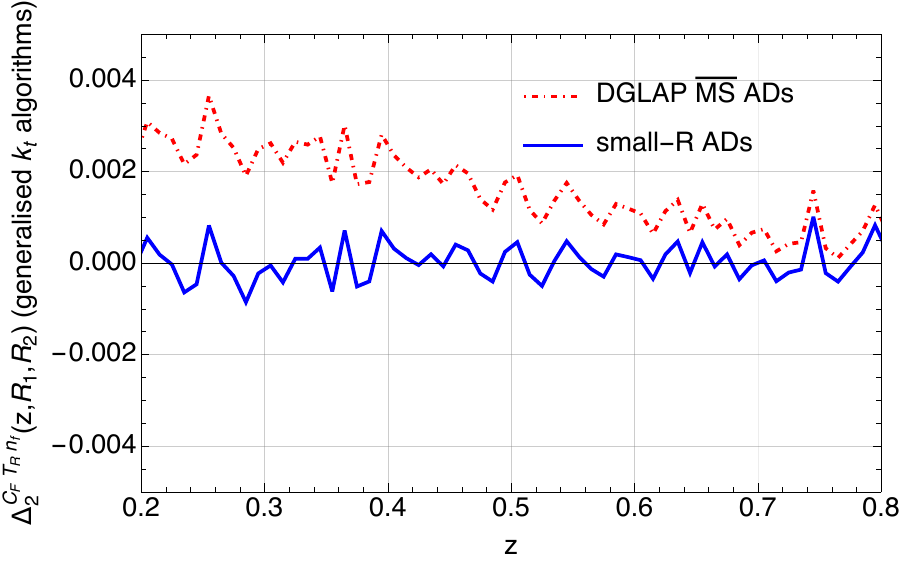}        
	\caption{Left: Function $\Delta_1(z,R_1)$ displaying the
          difference with the fixed-order prediction at
          ${\cal O}(\alpha_s)$. Right: Function $\Delta_2(z,R_1,R_2)$
          displaying the difference with the fixed-order prediction at
          ${\cal O}(\alpha^2_s)$ for the $C_F\,T_R\,n_f$ colour
          channel.}
	\label{fig:delta1}
\end{figure}

\begin{figure}[htbp]
        \centering
        \includegraphics[width=0.48\textwidth]{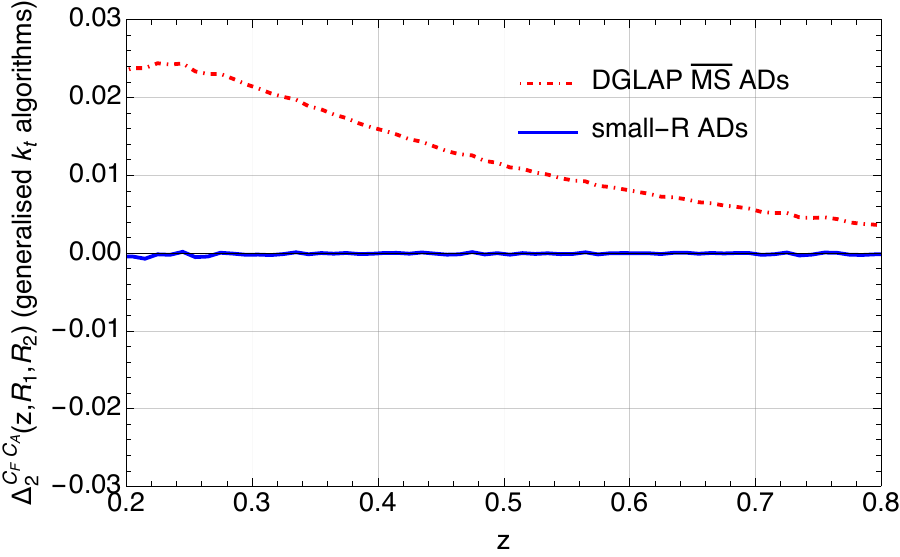}
	\includegraphics[width=0.48\textwidth]{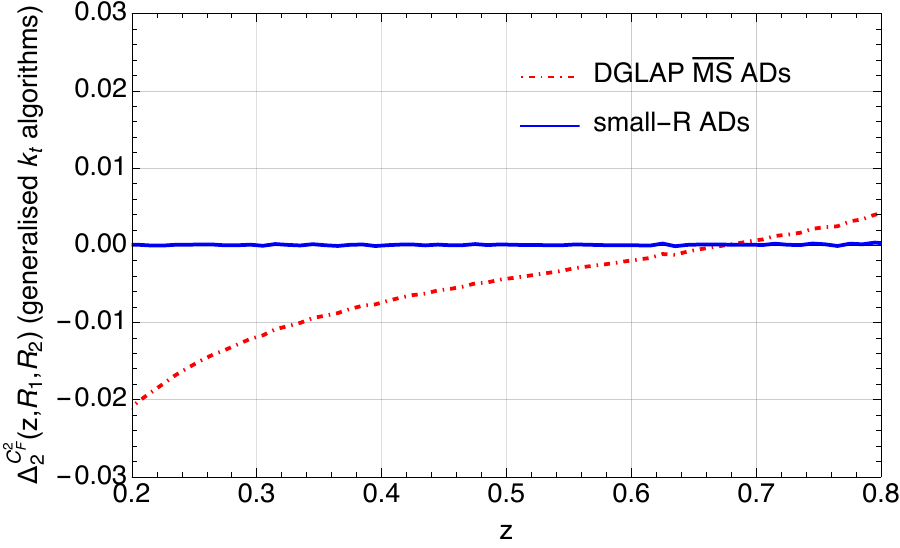}
	\caption{Left: Function $\Delta_2(z,R_1,R_2)$ displaying the
          difference with the fixed-order prediction at
          ${\cal O}(\alpha^2_s)$ for the $C_F\,C_A$ colour
          channel. Right: Function $\Delta_2(z,R_1,R_2)$ displaying the
          difference with the fixed-order prediction at
          ${\cal O}(\alpha^2_s)$ for the $C_F^2$ colour channel.}
	\label{fig:delta2}
\end{figure}

We start by considering the ${\cal O}(\alpha_s)$ expansion, and in
Fig.~\ref{fig:delta1} (left plot) we plot $\Delta_1(z,R_1)$ for
$R_1=0.01$. The small value of $R_1$ allows us to neglect subleading
power corrections in $R_1$ in the theoretical prediction. 
As expected, the result is consistent with zero within the statistical
fluctuations of \textsc{Event2}, in line with an NLL prediction for
the inclusive micro-jet spectrum.
We then move to ${\cal O}(\alpha_s^2)$. At this order, an NLL
prediction is expected to capture correctly all logarithmic terms
$\ln R^2$, but not the $R$-independent constant terms.
In this case, it is convenient to define a second quantity as
\begin{equation}\label{eq:delta2}
\Delta_2(z,R_1,R_2)\equiv \Delta_2(z,R_1)-\Delta_2(z,R_2)\,.
\end{equation}
Because of the difference between two jet radii, the quantity
$\Delta_2(z,R_1,R_2)$ does not contain the ${\cal O}(\alpha_s^2)$
constant terms at leading power in the limit $R^2 \ll 1$, which are
beyond our accuracy (i.e.~NNLL) in this study.
In the following, we show $\Delta_2(z,R_1,R_2)$ for $R_1=0.01$ and
$R_2=0.005$, separately for each of the colour structures, that is
$C_F^2$, $C_F\,T_R\,n_f$, $C_F\,C_A$. These are displayed in
Fig.~\ref{fig:delta1} (right plot) and Fig.~\ref{fig:delta2}. In these
figures, the red, dot-dashed line represents the prediction obtained
by using the standard NLO DGLAP anomalous dimensions in the
$\overline{\rm MS}$ scheme in Eq.~\eqref{eq:Djet}. On the other hand,
the blue, solid line represents our prediction, obtained by using the
$\hat{P}_{ik}$ kernels in the anomalous dimension matrix.
From the plots, we clearly observe that the DGLAP-like prediction for
$\Delta_2(z,R_1,R_2)$ is incompatible with zero. Instead, the new
anomalous dimensions obtained in this article lead to an excellent
agreement with the fixed order expectation within the statistical
fluctuations of \textsc{Event2} over a very wide range of $z$.
At the edges of the $z$ range we noticed a deviation between our
prediction from Eq.~\eqref{eq:small-R-FF} and \textsc{Event2}. This is
related to subleading-power terms in $R$ present in the latter
fixed-order prediction, which exhibit a divergent behaviour near the
endpoints $z=0$ and $z=1$. These are particularly pronounced in the
$C_F^2$ and $C_F\,C_A$ channels, where the splitting kernels are
divergent either at one of or both of the endpoints.
Eliminating this feature would require evaluating
$\Delta_2(z,R_1,R_2)$ at very small values of $R_1$ and $R_2$, which
due to numerical stability (as well as to the presence of technical
cutoffs) in $\textsc{Event2}$ is very challenging. For this reason we
cut the region near the extremities of the $z$ range in the plots.

\subsection{Anomalous dimensions for alternative jet algorithms: small-$y_{\rm cut}$ jets.}
\label{sec:small-ycut}
\begin{figure}[htbp]
        \centering
        \includegraphics[width=0.48\textwidth]{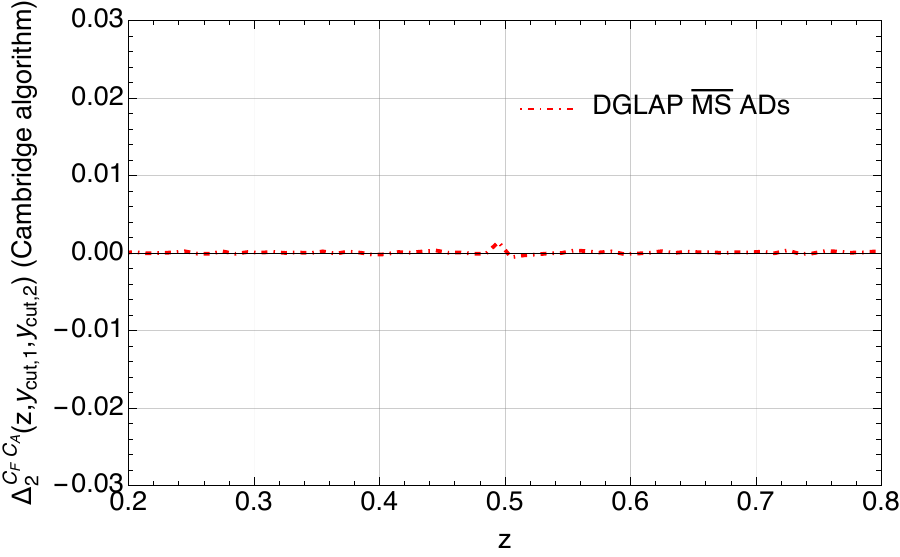}
	\includegraphics[width=0.48\textwidth]{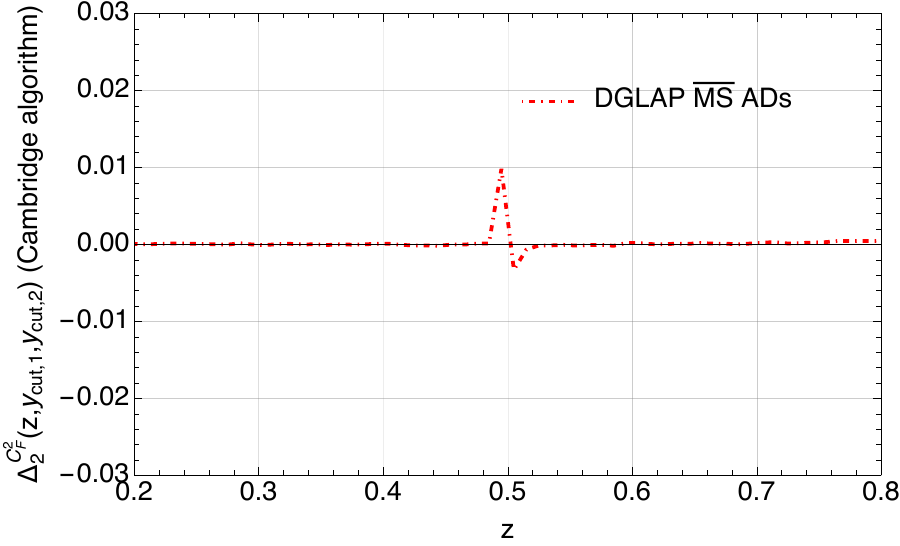}
        \includegraphics[width=0.48\textwidth]{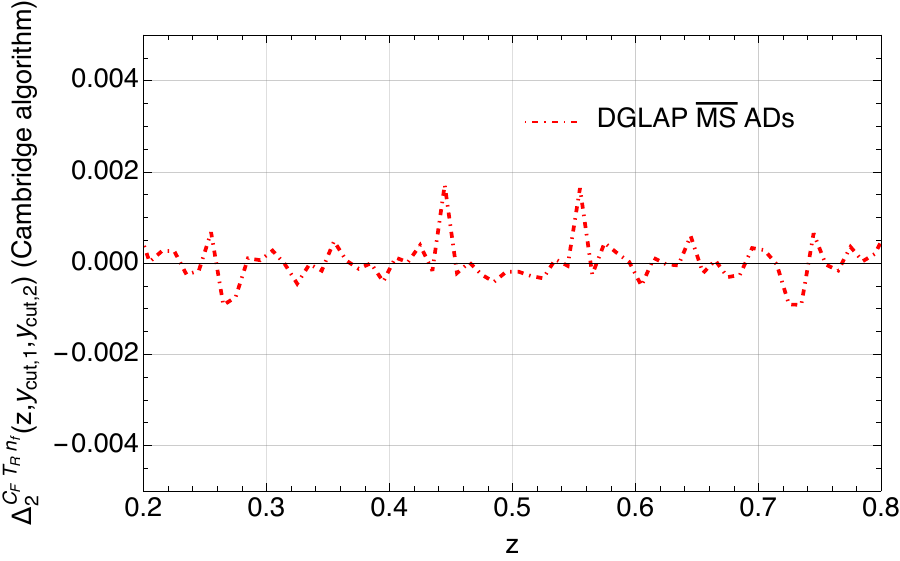}        

	\caption{Function $\Delta_2(z,y_{\rm cut,1}, y_{\rm cut,2})$
          for different colour factors.}
	\label{fig:delta2-ycut}
\end{figure}

It is instructive to investigate the dependence of the FF's anomalous
dimension on the kinematic cut applied on jets. Specifically, in
addition to the case of inclusive Cambridge-Aachen jets considered so
far, we also study Cambridge jets~\cite{Dokshitzer:1997in} with a
small $y_{\rm cut}$.
The FF in the latter case is defined by ordering proto-jets according
to their angular distance and then cluster them according to the $k_t$
measure. This means that two proto-jets $i$ and $j$ are recombined
into a single jet if
\begin{equation}
y_{ij} =
2\,\frac{\min\{E_i^2,E_j^2\}}{Q^2}\,\left(1-\cos\theta_{ij}\right) <
y_{\rm cut}\,,
\end{equation}
or else the less energetic of the two defines a jet and it is removed
from the proto-jets list.
This procedure amounts to setting a transverse momentum cut on the
final state jets, as opposed to an angular cut as in the small-$R$
case.
We repeated the calculation of Sec.~\ref{sec:FF} for this variant of
the FF and found that in this case the NLL result differs from the
small-$R$ one both at the level of the boundary conditions to the
evolution equation~\eqref{eq:Djet} as well as at the level of the
anomalous dimensions. Remarkably, we find that the anomalous
dimensions now coincide with the standard DGLAP ones in the
$\overline{\rm MS}$ scheme, while the boundary conditions are given
by:
\begin{align}
& D^{{\rm  jet}\,(0)}_q\left(z,E \sqrt{y_{\rm cut}}, E \sqrt{y_{\rm cut}} \right) = \\
 &
   \hspace{2.5cm}\frac{C_F}{2}\,\left(3-4\,z+z^2+4\,(1+z^2)\,\ln\frac{\min\{1-z,z\}}{1-z}\right)\, \left(\frac{1}{1-z}\right)_+\notag\\
  & \hspace{2.5cm} -
    \frac{C_F}{2\,z\,(1-z)}\,\bigg(4\,(2-3\,z+3\,z^2)\,\ln
    \,z \notag\\
    & \hspace{2.5cm}+(1-z)\,\left((5-z)\,z-4\,(2-2\,z+z^2)\,\ln\frac{\min\{1-z,z\}}{1-z}\right)\bigg)\notag\\
 & \hspace{2.5cm}-\frac{C_F}{6}\,\left(2\,\pi^2-3\,\left(7-6\,\ln
   2\right)\right)\,\delta(1-z) -2\, (\hat{P}_{qq}(z)+\hat{P}_{gq}(z))\,\ln 2\,,\notag
\end{align}
\begin{align}
& D^{{\rm  jet}\,(0)}_{\bar q}\left(z,E \sqrt{y_{\rm cut}},E \sqrt{y_{\rm cut}}\right) = D^{{\rm  jet}\,(0)}_q\left(z,E \sqrt{y_{\rm cut}},E \sqrt{y_{\rm cut}}\right) \,,\\
& D^{{\rm  jet}\,(0)}_g\left(z, E \sqrt{y_{\rm cut}}, E \sqrt{y_{\rm cut}}\right)
 =-4\,T_R\,n_f\,z\,(1-z)\notag\\
  & \hspace{2.5cm} -4\, \frac{C_A(1-z+z^2)^2+T_R\,n_f\,z\,(1-3\,z+4\,z^2-2\,z^3)}{(1-z)\,z}\,\ln\frac{(1-z)\,z}{\min\{1-z,z\}}\notag\\
  & \hspace{2.5cm} +\frac{1}{36}\left(C_A\,\left(131-12\,\pi^2-132\,\ln
    2\right)-2\,T_R\,n_f\,\left(17-24\,\ln 2\right)\right)\,\delta(1-z)\notag\\
  &\hspace{2.5cm} - \left(2\,\hat{P}_{gg}(z) + 4 \,n_f\,\hat{P}_{qg}(z)\right)\ln 2\,.
\end{align}
In Fig.~\ref{fig:delta2-ycut} we show the small-$y_{\rm cut}$
counterpart to the difference~\eqref{eq:delta2} between the
fixed-order calculation at ${\cal O}(\alpha_s^2)$ obtained with
\textsc{Event2} and our analytic prediction. We use
$y_{\rm cut,1} = 0.005^2$ and $y_{\rm cut,2}=0.001^2$ in our test, and
find perfect agreement between the two predictions.
These findings suggest that the DGLAP anomalous dimension in the
$\overline{\rm MS}$ scheme is in one-to-one correspondence with a
transverse momentum cut on final-state jets. This observation is
important in the context of reproducing DGLAP evolution beyond LL in
parton showers, which inevitably use a kinematic cutoff on the
generated radiation.

\subsubsection{A simple recipe to encode the small-$R$ evolution
  kernels into the DGLAP equation}
We can exploit the calculation performed in the previous section to
explore an idea similar in spirit to the one suggested in
Ref.~\cite{Dokshitzer:2005bf} to absorb the new terms in the two-loop
evolution kernels of the small-$R$ FF into a redefinition of the
evolution cutoff.
In particular, we consider modifying the collinear cutoff in the
small-$R$ resummation with a transverse-momentum scale $\mu=z\, E R$,
with $z$ being the longitudinal momentum fraction carried by the
jet. Specifically, the latter scale represents the maximum transverse
momentum carried by emissions within the jet.
We then evaluate the evolution equation~\eqref{eq:Djet} starting from
a low scale $\mu =z\, E R$ up to a high scale $\mu = E$. As a
consequence of this change of initial scale, the small-$R$ coefficient
functions~\eqref{eq:bc-D} are modified accordingly and the $\ln z$
terms in these equations get absorbed into the running of the
coupling.
In this case, we verified that the anomalous dimensions match the
DGLAP/small-$y_{\rm cut}$ ones, hence leading to a evolution
equation for these fragmentation functions at the NLL order that
coincides with the DGLAP equation.
It is important to stress that, in this prescription for the collinear
cut-off, the fraction $z$ is that of the final jet, and not that of an
intermediate state in the branching history implying that this
fraction does not enter the convolutions.  We have explicitly verified
that the prescription of using standard DGLAP evolution with the $z R$
dependent cut-off, agrees with our small-$R$ evolution equation at NLL
order.\footnote{A crucial identity needed for the verification is the
  following simple result for a chain of convolutions of identical
  splitting functions:
  $(\hat{P}_{ik}^{(0)}(z_1)\otimes\hat{P}_{ik}^{(0)}(z_2)\otimes\,...\,\otimes(\ln
  z_n \hat{P}_{ik}^{(0)}(z_n)))(z)=\frac{1}{n}\,\ln
  z\,(\hat{P}_{ik}^{(0)}(z_1)\otimes\hat{P}_{ik}^{(0)}(z_2)\otimes\,...\,\otimes\hat{P}_{ik}^{(0)}(z_n))(z)$.}
The above finding is a further confirmation of the observation made in
the previous section, of an intimate correspondence between the
$\overline{\rm MS}$ scheme and a transverse momentum cutoff.
While the $zR$ cut-off prescription restores a standard DGLAP form for
the evolution equation, it involves resumming logarithms of $z R$
rather than purely those in $R$. Hence while it works at NLL to
capture the logarithms of $R$, it also introduces potentially
uncontrolled logarithms of $z$ into the final result.

The prescription of starting the evolution at a scale proportional to
$z\,R$ concurs with the result of the formalism of
Ref.~\cite{Kang:2016mcy}, provided one also resums the logarithms of
$z$ consistently in the evolution equations. Specifically, in addition
to evolving the jet function between the scales $z\,R\,E$ and $z\,E$,
as done in Ref.~\cite{Kang:2016mcy}, it is crucial to evolve the hard
function between $E$ and $z\,E$ in order to get a full NLL resummation
of small-$R$ logarithms.
We note that this peculiar additional resummation of logarithms of $z$
necessary to bring the evolution equation into a DGLAP form would not
be necessary in the case of small-$y_{\rm cut}$ jets treated in the
previous section.

Furthermore, our result provides a crucial insight in the context of
reproducing DGLAP within a parton shower beyond LL. In this case, the
shower algorithm would operate with a cutoff related to the kinematics
of the intermediate state and to the evolution variable.
The simple recipe discussed in this section of stopping the evolution
at a scale proportional to $z\,R$ would then not be viable, in that
one does not have access to the final jet's $z$ fraction
during the showering process. We will present an in-depth discussion
of this point within a concrete shower algorithm in a forthcoming
publication.

\section{Conclusions}
\label{sec:conclusions}
With the eventual goal of testing a recent formulation of collinear
evolution using a generating functional
method~\cite{vanBeekveld:2023lsa} in mind, in this paper we have
analysed the fragmentation function of small-$R$ jets at NLL order,
via a fixed-order calculation at two loop order.
Here we performed for the first time a two loop calculation of the FF
in the presence of the angular cutoff set by the jet radius, and found
a difference in the two-loop anomalous dimension for small-$R$ FFs
relative to the standard DGLAP ones in the $\overline{\rm MS}$ scheme.
We have shown a calculation for the inclusive microjet spectrum at NLO
in the limit $R\ll 1$, focusing on the $C_F^2$ colour channel, which
is sufficient to infer the general form of the deviation from the
standard DGLAP anomalous dimensions in all colour channels.  This in
turn allowed us to determine the anomalous dimensions for small-$R$
jets at the two loop order. We confirmed our conclusions by comparing
to a numerical calculation from the fixed-order program
$\textsc{Event2}$ at ${\cal O}(\alpha_s^2)$.
An important first remark is that these findings are not just specific
to the Cambridge-Aachen algorithm but apply to all members of the
generalised $k_t$ family of algorithms, widely used at hadron
colliders. For the SISCone algorithm, the only difference at this
logarithmic order is encoded in the one-loop boundary condition of the
evolution equation~\eqref{eq:Djet} for the small-$R$ FF.
Finally, the result reported here agrees with the second-order
expansion of the generating functional given in
Ref.~\cite{vanBeekveld:2023lsa}, thereby validating this method for
the fragmentation of small-$R$ jets.
The derivation is outlined in Appendix~\ref{app:B2-derivation}.

Moving forward, it is interesting to observe that the simple form of
the difference between the small-$R$ and timelike DGLAP anomalous
dimensions amounts to terms of the same form as that responsible for
the violation of the Gribov-Lipatov reciprocity relation, which has
also been previously linked to a change of the kinematic
cutoff~\cite{Dokshitzer:2005bf}. To investigate what this implies in
the case of the FF we also performed a two loop calculation for an
alternative FF measured on Cambridge jets with a transverse momentum
cut ($y_{\rm cut}$), as well as for the small-$R$ FF with a cutoff
that represents the maximum possible transverse momentum of emissions
in the jet, proportional to $z \,R$. Remarkably, in these cases we
find that, besides a difference in the one-loop boundary conditions,
the two loop anomalous dimensions now coincide with the DGLAP ones in
the $\overline{\rm MS}$ scheme. This highlights a correspondence
between $\overline{\rm MS}$ and a transverse momentum cut, which is
critical to reproduce DGLAP evolution at higher orders with parton
showers.
Ultimately, it will also be important to assess the effect of our finding
in the context of the phenomenology of small-$R$ jets at the LHC and
future colliders, in view of the high precision that will be reached in
the study of the structure of hadronic jets.

\section*{Acknowledgments}
We thank Gavin Salam for valuable discussions during the course of
this project. This work has been partly funded by the European
Research Council (ERC) under the European Union's Horizon 2020
research and innovation program (grant agreement No 788223) (MvB, MD,
BKE, JH) and by the U.K.'s Science and Technologies Facilities Council
under grant ST/T001038 and ST/X00077X/1 (MD). The work of PM is funded
by the European Union (ERC, grant agreement No. 101044599, JANUS).
Views and opinions expressed are however those of the authors only and
do not necessarily reflect those of the European Union or the European
Research Council Executive Agency. Neither the European Union nor the
granting authority can be held responsible for them.
MD and PM would like to thank the Erwin-Schrödinger-Institute for
Mathematics and Physics, for their hospitality while part of this work
was being carried out. We also thank each other's institutions for
their hospitality at different stages of this project.
Finally, we would like to dedicate this article to the memory of Stefano Catani. 
Like for many topics in QCD, his seminal work underpins our results here, 
in particular through our direct use of the triple-collinear splitting functions to 
derive our main results.
\appendix

\section{Phase space parametrisation}
\label{app:phase-space}
In this appendix we provide the parametrisation of the three body
phase space used throughout the calculations in the article. In
$D=4-2\epsilon$ dimensions, the phase space $d\Phi_3$
reads
\begin{align}\label{eq:psnormal}
	\sd \Phi_3 =\frac{1}{\pi}  \frac{E^{4-4\epsilon}}{(4\pi)^{4-2\epsilon} \Gamma(1-2\epsilon)} \sd z_2 \sd z_3 \sd \theta_{13}^2 \sd  \theta_{23}^2 \sd \theta_{12}^2 (z_1z_2 z_3)^{1-2\epsilon} \Delta^{-1/2-\epsilon} \,\Theta(\Delta) \ \ ,
\end{align}
where the Gram determinant is given by
\begin{align}
	\Delta = 4 \theta_{ik}^2 \theta_{jk}^2 - \left(\theta_{ij}^2 - \theta_{ik}^2- \theta_{jk}^2\right)^2 ,\quad i\neq j \neq k\,,
\end{align}
and
\begin{equation}
\sum_{i=1}^3z_i=1\,.
\end{equation}

We can now introduce the parametrisation of Fig.~\ref{fig:b-splits},
for which the phase space measure becomes
\begin{align}\label{eq:psB}
  \sd \Phi_3^{(B)}\equiv\frac{x}{\pi}
  \frac{E^{4-4\epsilon}}{(4\pi)^{4-2\epsilon} \Gamma(1-2\epsilon)} \sd
  x \sd z_p \sd \theta_{13}^2 \sd  \theta_{23}^2 \sd \theta_{12}^2
  ((1-x) x^2 (1-z_p) z_p)^{1-2\epsilon} \Delta^{-1/2-\epsilon} \,\Theta(\Delta) \,.
\end{align}
Similarly, for the parametrisation of Fig.~\ref{fig:a-splits} we
obtain
\begin{align}
  \sd \Phi_3^{(A)}\equiv \frac{1-x}{\pi}
  \frac{E^{4-4\epsilon}}{(4\pi)^{4-2\epsilon} \Gamma(1-2\epsilon)} \sd
  x \sd z_p \sd \theta_{13}^2 \sd  \theta_{23}^2 \sd \theta_{12}^2
  ((1-x)^2 x (1-z_p) z_p)^{1-2\epsilon} \Delta^{-1/2-\epsilon}
  \,\Theta(\Delta) \,.
\end{align}

\section{Two loop anomalous dimensions for small-$R$ fragmentation}
\label{app:ADs}
In this appendix we report the explicit anomalous dimensions governing
the fragmentation of small-$R$ jets. Starting from the
equation~\eqref{eq:tildeP}, for quark fragmentation we obtain:
\begin{align}
 \delta\hat{ P}^{(1)}_{ qq}(z) &=
                              C_F^2\,\ln z\,\frac{1+z\,(4+z)+4\,(1+z^2)\,\ln(1-z)-(1
                              + 3 z^2) \,\ln z }{1-z}\notag\\
  & +2\, C_F\,T_R\,n_f \,\frac{13 \,(1-z^3) + 3 \ln z\,
    (4+3\,z\,(2+z)+3\,z\,(1+z)\,\ln z)}{9 z}\,,\\
\delta\hat{ P}^{(1)}_{ \bar{q}q}(z) &= 2\, C_F\,T_R\,n_f \,\frac{13 \,(1-z^3) + 3 \ln z\,
    (4+3\,z\,(2+z)+3\,z\,(1+z)\,\ln z)}{9 z} \,,
\end{align}
\begin{align}
\delta\hat{ P}^{(1)}_{ g q}(z) &=C_F^2 \frac{1}{3 z}
                              \bigg(3\,(5-z)\,(1-z)-2\,\pi^2\,\left(2-(2-z)\,z\right)+12\,(2-(2-z)\,z)\,{\rm
                              Li}_2(z)\\
&+ 3\ln z \,\big(4\,
  \left(2-(2-z)\,z\right)\,\ln(1-z)+z\,\left(6-2\,z+(2-z)\,\ln
                                                 z\right)\big)
                                                 \bigg)\notag\\
  & -4\, C_F\,T_R\,n_f \frac{\left(2-(2-z)\,z\right)\,\ln z}{3
    z}\notag\\
  & + C_F\,C_A\,\frac{1}{9 z}\bigg(6 \,\pi^2 \left(2-(2-z)\,z\right)
    -(1-z)\,\left(71+z\,(17+26\,z)\right)\notag\\
  & +3\,\ln z\,\left(-22-5 \,z\,(2-z)-12\,(1+z+z^2)\,\ln z\right)-36\,
    \left(2-(2-z)\,z\right)\,{\rm Li}_2(z)\bigg)\,.\notag
\end{align}
Similarly, for gluon fragmentation we obtain:
\begin{align}
  \delta\hat{ P}^{(1)}_{ gg}(z) &= -C_A^2\,\frac{2\,\ln z}{3
                               \,(1-z)\,z}\bigg(11
                               -18\,z+3\,z^2-18\,z^3+11\,z^4-12\,(1-z+z^2)^2\,\ln(1-z)\notag\\
  &+6\,(1+3\,z^2-4\,z^3+z^4)\,\ln z\bigg) -
    8\,C_A\,T_R\,n_f\,\frac{(1-z+z^2)^2\,\ln z}{3\,(1-z)\,z}\notag\\
  & +
    C_F\,T_R\,n_f\,\left(-\frac{52}{9}\frac{1-z^3}{z}-\frac{4}{3}\left(3+6\,z+4\,z^2\right)\,\ln
    z+4\,(1+z)\,\ln^2z\right)\,,
\end{align}
\begin{align}
  \delta\hat{ P}^{(1)}_{ qg}(z) &=\delta\hat{ P}^{(1)}_{ \bar{q}g}(z) =
                               C_F\,T_R\,n_f\,\bigg(1-6\,z+5\,z^2+\frac{2}{3}\,\pi^2\,(1-2\,z+2\,z^2)+(1-2\,z)\,\ln
                               z\notag\\
  & -\,(1-2\,z+4\,z^2)\,\ln^2 z-(4-8\,z+8\,z^2)\,{\rm
    Li}_2(z)\bigg)\notag\\
                             &+C_A\,T_R\,n_f\,\frac{1}{9}\bigg(-9-6\,\pi^2+\frac{26}{z}+54\,z+12\,\pi^2\,z-71\,z^2-12\,\pi^2\,z^2\notag\\
  &+\frac{6\,\ln
    z}{z}\left(4+6\,z+18\,z^2-9\,z^3+6\,z\,(1-2\,z+2\,z^2)\,\ln(1-z)\right)\notag\\
  & + 18\,(1+4\,z)\,\ln^2z+36\,\left(1-2\,z+2\,z^2\right)\,{\rm
    Li}_2(z)\bigg)\,.
\end{align}

\section{Derivation of two loop anomalous dimensions from $B_2^q(z)$}
\label{app:B2-derivation}
In this appendix we discuss the derivation of the two loop FF for
small-$R$ jets from the generating functionals
formalism~\cite{Konishi:1979cb,Bassetto:1983mvz,Dokshitzer:1991wu,Ellis:1996mzs,Dasgupta:2014yra}
extended to NLL in Ref.~\cite{vanBeekveld:2023lsa}.
This derivation summarises the original calculation of the small-$R$
anomalous dimension, which led us to uncover the discrepancy with the
timelike DGLAP case. It also serves as an important test of the method
of Ref.~\cite{vanBeekveld:2023lsa}.

We work in the non-singlet (NS) case to simplify the notation, but
analogous considerations hold for the other flavour channels.
Our starting point is the definition of the NS FF in terms of the
quark generating functional $G_q(x,t)$, where $t$ is the evolution
time (cf. Eq.~(2.1) of Ref.~\cite{vanBeekveld:2023lsa}). This is
related to a resolution scale (angle) $\mu = E\,\theta$ by
\begin{equation}\label{eq:t}
t = \int_{\mu^2}^{E^2}\frac{d
  \mu'^2}{\mu'^2}\frac{\alpha_s(\mu'^2)}{2\pi}\,,
\end{equation}
where $E=Q/2$ is the energy of the initial fragmenting parton (quark
in the NS case).
The FF can be obtained by taking the derivatives of the quark
generating functional $G_q$ w.r.t.~the probing
function~\cite{Konishi:1979cb,Bassetto:1983mvz,Dokshitzer:1991wu}
(source) $u$ which has the function of tagging a final state parton.

We start with LL for the sake of simplicity, and then discuss the NLL
case. The evolution of $G_q$ with the evolution time $t$ is driven by
the integral equation
\begin{align}\label{eq:GF-LL}
  G_q(x,t) &= u\, \Delta_q(t) + \int_{t}^{t_{0}}\ d t' \int_{z_0}^{1-z_0} d z\,
             P_{qq}(z)\,G_q(x\,z,t') \,G_g(x\,(1-z),t') \frac{\Delta_q(t)}{\Delta_q(t')}\,,
\end{align}
where $ P_{qq}(z) = C_F (1+z^2)/(1-z)$ denotes the unregularised
$q\to q g$ LO splitting function and $\Delta_q(t)$ is the Sudakov form
factor defined as~\footnote{cf. Eq.~(2.5) of
  Ref.~\cite{vanBeekveld:2023lsa}. The limits of the collinear
  ($t_0\to \infty$) and IR ($z_0\to 0$) cutoffs are meant to be taken
  at the level of physical observables. The dependence on the cutoffs
  for IRC safe observables will cancel against that in the real
  corrections up to power corrections, made negligible by taking a
  small numerical cutoff. The latter vanish in the calculation
  reported here since the limits are taken analytically.}
\begin{align}\label{eq:sud-quark}
\ln \Delta_q(t) &= - \int_t^{t_0} d t' \int_{z_0}^{1-z_0}d z\, P_{qq}(z)\,.
\end{align}

We now proceed to calculate the NS FF. A first simplification comes
from observing that the gluons produced in the branching of the quark
do not fragment further if one considers the NS channel. For this
reason we can approximate $G_g$ with its first order expansion
\begin{equation}
G_g = u + {\cal O}(\alpha_s)\,,
\end{equation}
hence neglecting any radiative corrections. This implies that the
evolution of $G_q$ in the NS channel simply amounts to a sequence of
angular-ordered primary gluon emissions off the fragmenting quark.
The small-$R$ NS FF at LL is then given by
\begin{equation}\label{eq:DNSR}
  D_{\rm NS}^{\rm jet}(z,\mu, E R) = \sum_n \int d P_n  
  \delta(z-\prod_{i=1}^{n-1}
  z_i)\prod_{i=1}^{n-1}\Theta(\theta_i^2-R^2) = \sum_{n} D_{\rm
    NS}^{\rm jet, (n)}(z,\mu, E R)\,.
\end{equation}
The \textit{emission probabilities} $d P_n$ are calculated with the
formula
\begin{equation}
\int d P_n \equiv \left. \frac{1}{n!} \frac{\delta ^n}{\delta u^n}
  G_q\right |_{u=0}\,.
\end{equation}
We define the functional derivative by the above equation to
effectively act as an ordinary derivative, whereas kinematic
phase-space constraints (e.g. the observable's measurement function)
are explicitly added for each $d P_n$ (see Eq.~\eqref{eq:DNSR} for the
FF case).\footnote{An alternative definition, albeit practically
  equivalent, of the functional derivative is given in
  Ref.~\cite{Banfi:2021xzn}.}
The delta function in Eq.~\eqref{eq:DNSR} fixes the longitudinal
momentum of the final state quark to be $z$. This is easily obtained
by noticing that each gluon carries a relative fraction $1-z_i$ of the
momentum of the parent, which sets the energy of the final state quark
to $E$ times the product of all $z_i$ fractions.
Finally, the theta function in Eq.~\eqref{eq:DNSR}, with $\theta_i$
being the angle of gluon $i$ w.r.t.~the final state
quark~\footnote{Due to strong angular ordering, this coincides with
  the angle w.r.t.~the emitter.}, implements the Cambridge clustering
condition. This ensures that we consider only gluons which are not
recombined with the quark jet, and hence change the jet's momentum
fraction.
In terms of the evolution time $t$, this constraint simply
translates to
\begin{equation}
t_i < t_R \equiv \int_{E^2 R^2}^{E^2}\frac{d \mu'^2}{\mu'^2}\frac{\alpha_s(\mu'^2)}{2\pi}\,.
\end{equation}
This effectively replaces the collinear cutoff $t_0$ in
Eq.~\eqref{eq:GF-LL}, including in the Sudakov form factor. Similarly,
we also take the limit of the IR cutoff $z_0\to 0$ given the IR safety
of the FF that we are computing.
From Eq.~\eqref{eq:DNSR} we obtain
\begin{align}\label{eq:DNSR-boundary}
D_{\rm  NS}^{\rm jet, (1)}(z,\mu, E R) &= \Delta_q(t)\,\delta(z-1)\,,\\
D_{\rm  NS}^{\rm jet, (2)}(z,\mu, E R) &= \Delta_q(t)
  \int_{t}^{t_R} d t_1\int_0^1 d z_1 P_{qq}(z_1) \,\delta(z-z_1)\,,\notag\\
D_{\rm  NS}^{\rm jet, (3)}(z,\mu, E R) &= \Delta_q(t)
  \int_{t}^{t_R} d t_1 \int_{t_1}^{t_R} d t_2\int_0^1 d z_1\,d z_2 P_{qq}(z_1) P_{qq}(z_2)\, \delta(z-z_1z_2)\,,\notag\\
  ... &= ...\,,\notag\\
D_{\rm  NS}^{\rm jet, (n)}(z,\mu, E R) &= \Delta_q(t)
  \int_{t}^{t_R} d t_1\,...\,\int_{t_{n-2}}^{t_R} d t_{n-1}\int_0^1
  \left(\prod_{i=1}^{n-1}d z_i P_{qq}(z_i)\right) \,\delta(z-\prod_{i=1}^{n-1}z_i)\,.\notag
\end{align}
The scale $t$ in the above equations is related by Eq.~\eqref{eq:t} to
the upper bound on the angle at which the evolution is stopped. In
order for all logarithmic terms to be resummed in the FF we set the
final scale to $\mu=E$ corresponding to the final $t=0$.
We further introduce the quantity
\begin{equation}
\Sigma(z,t) \equiv \sum_\ell \int_{t}^{t_R} d t_1\,...\,\int_{t_{\ell-1}}^{t_R} d t_{\ell}\int_0^1
  \left(\prod_{i=1}^{\ell}d z_i P_{qq}(z_i)\right) \,\delta(z-\prod_{i=1}^{\ell}z_i)\,,
\end{equation}
which allows us to write the NS FF at LL as
\begin{equation}~\label{eq:DNS-LL}
  D_{\rm  NS}^{\rm jet}(z,\mu, E R) = \Delta_q(t) \left(\delta(z-1) + \Sigma(z,t)\right)\,.
\end{equation}

What we are interested in here is the evolution of
$D_{\rm NS}^{\rm jet}(z,\mu, E R)$ w.r.t.~the resolution scale
$\mu$. This is related to the evolution in $t$ by
\begin{equation}\label{eq:diff}
\frac{d}{d\ln \mu^2} = - \frac{\alpha_s(\mu^2)}{2\pi}\,\frac{d}{d t}\,.
\end{equation}
We can then obtain the evolution equation for the FF by acting with
the above derivative on $D_{\rm NS}^{\rm jet}(z,\mu, E R)$.
We arrive at
\begin{align}\label{eq:evo-LL}
 \frac{ d D_{\rm  NS}^{\rm jet}(z,\mu, E R)}{d\ln\mu^2} &=
 \frac{\alpha_s(\mu^2)}{2\pi} \int_0^1 d y \,P_{qq}(y) \bigg[ \frac{\Delta_q(t)}{y}\left(\delta(z/y-1)+\Sigma(z/y,t)\right)\\
                                                        & - \Delta_q(t) \left(\delta(z-1) + \Sigma(z,t)\right)\bigg] 
= \frac{\alpha_s(\mu^2)}{2\pi} \int_z^1 \frac{dy}{y}\left(P_{qq}(y) \right)_+D_{\rm  NS}^{\rm jet}\left(\frac{z}{y},\mu, E R\right)\,.\notag
\end{align}
The first term in the above equation arises from the derivative of
$\Sigma$, while the second from the derivative of the Sudakov
$\Delta_q$. The LL evolution equation in Eq.~\eqref{eq:evo-LL} agrees
with the DGLAP equation, upon noticing that the plus prescription
acting on the whole unregularised splitting function $P_{qq}(y)$ is
fully equivalent to the standard regularised splitting function
$\hat{P}_{qq}(y)$.

As a next step, we now derive the corresponding equation at NLL. The
starting point is the NLL evolution equation which
reads~\cite{vanBeekveld:2023lsa} 
\begin{align}\label{eq:Kvquark}
 G_q(x,t)= u\, \Delta_q(t,x) \,+&
                               \int_t^{t_0}d t'\int_{z_0}^{1-z_0}d z\, \,G_q(x\,z,t')
  \,G_g(x\,(1-z),t')\frac{\Delta_q(t,x)}{\Delta_q(t',x)}\,{\mathcal P}_{q}(z,t',x)\notag\\
&+ {\mathbb K}_q^{\rm finite}[G_q,G_g] \,,
\end{align}
where
\begin{equation}\label{eq:Pq}
{\mathcal P}_{q}(z,t',x) \equiv {\mathcal
  P}_{q}(z,t')-\frac{\alpha_s(\mu'^2)}{2\pi} P_{qq}(z)\,b_0\,\ln x^2\,.
\end{equation}
Here $x$ is the longitudinal momentum fraction of the quark branching
at the angular scale $t'$. The term proportional to $\ln x^2$ in the
r.h.s.~of Eq.~\eqref{eq:Pq} has the role of ensuring that the
effective scale of the strong coupling is the energy of the parton
that is branching multiplied by the angular scale of the
branching.\footnote{In the case of ${\mathcal P}_{q}$,
    we have replaced the argument $\theta$ with $t$ to ease the
    notation. The two quantities are related by
    Eq.~\eqref{eq:t}. Moreover, we have explicitly expanded out the
    term proportional to $\ln x^2$ from the definition of the
    evolution time w.r.t.~the notation of
    Ref.~\cite{vanBeekveld:2023lsa} (cf. Eq.~(2.1) there), as this
    plays a central role in the derivation shown
    here.}
The quantities ${\mathcal P}_{q}(z,t)$ and
${\mathbb K}_q^{\rm finite}$ are defined in Eq.~(2.10) and Appendix C
of Ref.~\cite{vanBeekveld:2023lsa}, respectively, and they are derived
from the exact $1\to 3$ splitting functions and corresponding virtual
corrections.
Accordingly, the LL Sudakov form factor $\Delta_q(t)$ is also upgraded
to NLL and defined as
\begin{align}\label{eq:sud-quark}
\ln \Delta_q(t,x) &= - \int_t^{t_0} d t' \int_{z_0}^{1-z_0}d z\, {\mathcal P}_{q}(z,t',x)\,.
\end{align}
A few remarks are in order. The quantity ${\mathcal P}_{q}(z,t)$,
referred to as the inclusive emission probability, encodes the
next-to-leading order cross section for the radiation of a gluon of
momentum $1-z$ differentially both in $z$ and in the angle between the
gluon and the emitting quark. This is encoded in the anomalous
dimension ${\cal B}_2^q(z)$ obtained in Ref.~\cite{Dasgupta:2021hbh}.
The additional term ${\mathbb K}_q^{\rm finite}$ encodes the
fully-differential structure of the $1\to 3$ splitting functions,
which corrects the inclusive approximation made in
${\mathcal P}_{q}(z,t)$. This guarantees a full coverage of the
$1\to 3$ splitting phase space and corresponding splitting functions.

An important comment concerns the clustering condition as implemented
in Eq.~\eqref{eq:DNSR}. As discussed in the main text, in the
small-$R$ limit one can neglect any clustering of two or more
emissions, in that they would result in a power-suppressed term, as
well as the recoil of the jet axis. For this reason, the clustering
condition in Eq.~\eqref{eq:DNSR} (with $\theta_i$ denoting the angle
w.r.t.~the final state quark) remains valid at NLL.
From here we essentially follow the same procedure used in the LL
case, with more tedious calculations due to the presence of the extra
term ${\mathbb K}_q^{\rm finite}$. An important difference
w.r.t.~Eq.~\eqref{eq:DNS-LL} is that, at NLL, the FF receives a
contribution from ${\cal O}(\alpha_s)$ non-logarithmic terms that are
obtained by matching the GFs prediction at ${\cal O}(\alpha_s)$ to the
calculation of the small-$R$ FF at the same order.
This leads to the following expression for the NLL FF in the NS
channel
\begin{equation}\label{eq:FF-NLL-GF}
  D_{\rm  NS}^{\rm jet}(z,\mu, E R) =
  \left(1+\frac{\alpha_s(E^2)}{2\pi}\,D^{{\rm jet}\,(0)}_{q,\,{\rm
        NS}}\left(z,E R, E R\right)\right)\otimes \Delta_q(t) \left(\delta(z-1) + \Sigma(z,t)\right)\,,
\end{equation}
where $\Delta_q(t)\equiv \Delta_q(t,1)$ and
$D^{{\rm jet}\,(0)}_{q,\,{\rm NS}}\left(z,E R, E R\right)$ is the
non-singlet part of the quantity defined in Eq.~\eqref{eq:bc-D}.
It reads
\begin{align}\label{eq:bc-D-NS}
D^{{\rm  jet}\,(0)}_{q,\,{\rm NS}}\left(z,E\,R, E\,R\right)
  &=-2\,C_F\,(1+z^2)\,\left(\frac{\ln(1-z)}{1-z}\right)_+ - C_F\,\left((1-z)+2\frac{1+z^2}{1-z}\,\ln
    z\right)\notag\\
  &+C_F\,\left(\frac{13}{2}-\frac{2}{3}\,\pi^2\right)\,\delta(1-z)\,.
\end{align}
The second term to the right of the convolution sign in
Eq.~\eqref{eq:FF-NLL-GF} is the result of the functional derivatives
of the quark GF, where the quantity $\Sigma(z,t)$ at NLL is now given
by
\begin{align}\label{eq:sigmaNLL}
\Sigma(z,&t)\equiv \sum_\ell \int_{t}^{t_R} d t_1 \int_0^1 d z_1
\,\left({\cal
            P}_q(z_1,t_1,1)+\chi(z_1,t_1)\right)\frac{\Delta_q(t_1,z_1)}{\Delta_q(t_1,1)}\,...\,\\&\times\int_{t_{\ell-1}}^{t_R}
  d t_{\ell}\int_0^1 d z_\ell \,\left({\cal P}_q(z_\ell,t_\ell, z_1\,z_2\,...\,z_{\ell-1})+\chi(z_\ell,t_\ell)\right) \frac{\Delta_q(t_\ell, z_1\,z_2\,...\,z_{\ell})}{\Delta_q(t_\ell,z_1\,z_2\,...\,z_{\ell-1})}
\,\delta(z-\prod_{i=1}^{\ell}z_i)\,.\notag
\end{align}
The new kernel $\chi$ arises from the functional derivatives of
${\mathbb K}_q^{\rm finite}$ and, using the notation of the main text,
it is given by~\footnote{Eq.~\eqref{eq:chi} can be evaluated using the
  relation
  $\int \frac{d\theta'^2}{\theta'^2}\delta(t-t') = \int
  \frac{d\theta'^2}{\theta'^2}
  \frac{2\pi}{\alpha_s}\theta^2\delta(\theta^2-\theta'^2)$.}
\begin{align}\label{eq:chi}
\chi(z,t) &\equiv \frac{1}{2!} \int
  \left.d\Phi_3^{(A)} \frac{(8\pi\,\alpha_s(E^2\,\theta_{12,3}^2))^2}{s_{123}^2}\langle P\rangle_{C_F\,(C_F-C_A/2)}\right|_{\epsilon=0}\\&\qquad\qquad\qquad\times \delta(t - t_{12,3})\,\delta(z-z_p (1-x))\Theta(\theta_{12,3}^2-R^2)\notag\\
& + \int \left. d\Phi_{3}^{(B)} (8 \pi\,\alpha_s(E^2\,\theta_{13}^2))^2\left(\frac{1}{s_{123}^2}\,\langle
                        P\rangle_{C_F^2} - \frac{ {\cal J}(x,z_p)}{E^4}\,\frac{P^{(0)}_{qq}(x,\epsilon)}{\theta_{13}^2}
    \frac{P^{(0)}_{qq}(z_p,\epsilon)}{\theta_{23}^2}\right)\right|_{\epsilon=0} \notag\\ &\qquad\qquad\qquad\times\delta(t-t_{1,3}) \left(\delta(z-z_p x)-\delta(z-x)\right) \Theta(\theta_{13}^2-R^2) \Theta(\theta_{13}^2-\theta_{23}^2)\,,\notag
\end{align}
with ${\cal J}(x,z_p)$ given in Eq.~\eqref{eq:JPS}.
To evaluate the derivative w.r.t.~$\ln\mu^2$, we make use of
the following equation
\begin{align}\label{eq:derivative}
&  \frac{d\, \Sigma(z,t)}{d t} \simeq - \int_z^1 \frac{d y}{y} \,\left( {\mathcal  P}_{q}(y,t)+\chi(y,t)\right)
  \left(\delta\left(\frac{z}{y}-1\right)+\Sigma\left(\frac{z}{y},t\right)\right)\\ &+
 2\,b_0\, 
 \int_{E^2R^2}^{\mu^2}\!\!\frac{d\mu'^2}{\mu'^2}\frac{\alpha^2_s(\mu'^2)}{(2\pi)^2}\int_z^1 \frac{d y}{y}\ln y\,P_{qq}(y)\int_{z/y}^1\frac{d
 y'}{y'} \left(P_{qq}(y')\right)_+\left(\delta\left(\frac{z}{yy'}-1\right)+\Sigma\left(\frac{z}{yy'},t\right)\right)\,.\notag
\end{align}
In Eq.~\eqref{eq:derivative} we have exploited the fact that, at NLL,
one only needs to retain a single insertion of the $\ln x^2$ term in
Eq.~\eqref{eq:Pq}, while one can neglect subleading terms stemming
from multiple insertions of this contribution to ${\cal P}_q(z,t,x)$.
Taking the derivative of Eq.~\eqref{eq:FF-NLL-GF} using
Eq.~\eqref{eq:diff} and \eqref{eq:derivative} we obtain the following
evolution equation
\begin{align}\label{eq:evo-NLL}
 &\frac{ d D_{\rm  NS}^{\rm jet}(z,\mu, E R)}{d\ln\mu^2}
  =\frac{\alpha_s(\mu^2)}{2\pi} \int_z^1 \frac{d y}{y}
  \left[\left({\cal P}_q(y,t) \right)_++\chi(y,t)\right]D_{\rm
  NS}^{\rm jet}\left(\frac{z}{y},\mu, E R\right)\,\\
  &-b_0 \int_{E^2R^2}^{\mu^2}\frac{d\mu'^2}{\mu'^2}\frac{\alpha^2_s(\mu'^2)}{(2\pi)^2}\frac{\alpha_s(\mu^2)}{2\pi} \int_z^1\frac{dy}{y} \int_{z/y}^1\frac{dy'}{y'} 2\ln y\,P_{qq}(y) \left(P_{qq}(y')\right)_+ D_{\rm  NS}^{\rm jet}\left(\frac{z}{yy'},\mu, E R\right)\,\notag\\
& \equiv \frac{\alpha_s(\mu^2)}{2\pi} \left[\left({\cal P}_q(z,t) \right)_++\chi(z,t)-b_0 \int_{E^2R^2}^{\mu^2}\frac{d\mu'^2}{\mu'^2}\frac{\alpha^2_s(\mu'^2)}{(2\pi)^2}\delta\hat{P}_{qq}^{(1)}(z)\right]\otimes D_{\rm  NS}^{\rm jet}(z,\mu, E R)\,,\notag
\end{align}
with $\delta\hat{P}_{qq}^{(1)}(z)$ as defined in
Eq.~\eqref{eq:result}.
We now comment on the evolution kernel of r.h.s. of
Eq.~\eqref{eq:evo-NLL}, and its relationship to the small-$R$
anomalous dimension derived in the main text.
A first comment concerns the absence of the colour channels $C_F C_A$
and $C_F T_R n_f$ in Eq.~\eqref{eq:chi}. Let us examine the
contribution of the corresponding $1\to 3$ splitting functions to the
NS channel.
We start by reminding the reader that these channels contribute to the
inclusive emission probability ${\cal P}_q(z,t)$ via the integral of
the corresponding $1\to 3$ splitting functions with the longitudinal
momentum of the quark after the first splitting fixed. For the
$C_F C_A$ and $C_F T_R n_f$ channels, this corresponds to fixing the
longitudinal momentum fraction of the final quark with the same
flavour as the one that initiated the fragmentation. This can be
intuitively understood by looking at Fig.~\ref{fig:a-splits} for the
$C_F T_R n_f$ channel, where the momentum fraction $z_3$ is kept fixed
(and analogously for the $C_F C_A$ term).
On the other hand, the function $\chi(z,t)$ arises from the functional
derivatives of ${\mathbb K}_q^{\rm finite}$, which is sensitive to the
differential structure to the $1\to 3$ splitting. This acts as a
correction to the inclusive approximation made in the definition of
${\cal P}_q(z,t)$.
Since in the NS channel we tag the final state quark, no correction
from ${\mathbb K}_q^{\rm finite}$ survives if the final state quark
coincides with the quark after the first splitting, as it is the case
for the $C_F C_A$ and $C_F T_R n_f$ contributions. As a consequence,
the only contribution from these colour factors is encoded in
${\cal P}_q(z,t)$.

A second important comment about Eq.~\eqref{eq:evo-NLL} is the
resummation scheme~\cite{Catani:2000vq} used in the evolution of the
generating functionals. As discussed in
Ref.~\cite{vanBeekveld:2023lsa}, Eq.~\eqref{eq:evo-NLL} is defined in
a resummation scheme in which all non-logarithmic terms, captured in
the matching coefficient are evaluated at the hard scale $\mu\sim
E$. This is reflected in Eq.~\eqref{eq:FF-NLL-GF}, where the
non-logarithmic terms
$ D^{{\rm jet}\,(0)}_{q,\,{\rm NS}}\left(z,E R, E R\right)$ are
evaluated at the hard scale.
This implies that the logarithmic terms that would be otherwise
generated by the running of the low-energy boundary conditions to the
evolution equation~\eqref{eq:Djet} are already encoded in the kernel
of Eq.~\eqref{eq:evo-NLL} rather than in the boundary
conditions. Specifically, this means that the for the solutions to
Eq.~\eqref{eq:evo-NLL} and Eq.~\eqref{eq:Djet} (in the NS channel) to
be equivalent, we must verify that
\begin{align}\label{eq:conjecture}
\frac{\alpha_s(\mu^2)}{2\pi} \left[\left({\cal P}_q(z,t)
  \right)_++\chi(z,t)\right] &= \frac{\alpha_s(\mu^2)}{2\pi}
\hat{P}^{(0)}_{qq}\left(z\right) \notag\\&+\frac{\alpha^2_s(\mu^2)}{(2\pi)^2}
\left(\hat{P}^{(1) ,\, {\rm NS}}_{qq}\left(z\right) +b_0\, D^{{\rm
  jet}\,(0)}_{q,\,{\rm NS}}\left(z,E R, E R\right)\right)\,,
\end{align}
where $b_0= 11/6 \,C_A - 2/3\,T_R\,n_f$.%
\footnote{Note that the terms of Eq.~\eqref{eq:evo-NLL} and
  Eq.~\eqref{eq:Djet} proportional to $\delta \hat{P}^{(1)}_{qq}(z)$
  are trivially in agreement at NLL after integrating over $\mu'$
  (c.f.~Eq.~\eqref{eq:coupling-to-NSL}).}
Since we are working with the NS channel, now $\hat{P}^{(1) ,\, {\rm NS}}_{qq}$
only contains the NS contributions, i.e.
\begin{equation}\label{eq:PNS}
\hat{P}^{(1),\,{\rm NS}}_{qq}(z) \equiv \hat{P}^{(1),\, {\rm V}}_{qq}(z) - \delta \hat{P}^{(1)}_{qq}(z)\,,
\end{equation}
where $\hat{P}^{(1),\, {\rm V}}_{qq}(z)$ denotes the standard
regularised time-like splitting function in the NS channel (see
e.g. in Chapter 6 of Ref.~\cite{Ellis:1996mzs}).
While this correspondence is trivially true at ${\cal O}(\alpha_s)$,
in order to establish it at ${\cal O}(\alpha_s^2)$ we have verified
Eq.~\eqref{eq:conjecture} numerically for $z\neq 1$, finding perfect
agreement.
To analyse the case $z=1$, we can simply check that
Eq.~\eqref{eq:conjecture} holds at the integral level
(i.e. integrating over $z\in [0,1]$). The integral over $z$ of the
l.h.s. amounts to the integral of the first term in Eq.~\eqref{eq:chi}
as everything else vanishes upon integration either because of the
plus prescription (for what concerns
$\left({\cal P}_q(z,t) \right)_+$) or due to the difference of
$\delta$ functions (for the second term in Eq.~\eqref{eq:chi}). This
gives
\begin{equation}
\int_0^1 d z \frac{\alpha_s(\mu^2)}{2\pi} \chi(z,t) =
\frac{\alpha^2_s(\mu^2)}{(2\pi)^2} C_F\left(C_F-\frac{C_A}{2}\right)\frac{13-2\,\pi^2+8\,\zeta_3}{4}\,,
\end{equation}
in agreement with the integral of the NS regularised splitting
function, as predicted by Eq.~\eqref{eq:conjecture}.
We can now simplify further the structure of
  Eq.~\eqref{eq:evo-NLL} by writing, at NLL, the integral over $\mu'$
  as
  \begin{equation}\label{eq:coupling-to-NSL}
-  b_0  \int_{E^2R^2}^{\mu^2}\frac{d\mu'^2}{\mu'^2}\frac{\alpha^2_s(\mu'^2)}{(2\pi)^2}
    \simeq \frac{\alpha_s(\mu^2)}{2\pi}-\frac{\alpha_s(E^2R^2)}{2\pi}\,.
  \end{equation}
  It is now straightforward to see that the last term in the r.h.s. of
  Eq.~\eqref{eq:evo-NLL} has the role of changing the scale of
  $\delta\hat{P}_{qq}^{(1)}(z)$ in Eq.~\eqref{eq:conjecture} from
  $\mu^2$ to $E^2 R^2$. For consistency with the notation in the main
  text we also change the scale of the coupling multiplying the
  boundary condition
  $D^{{\rm jet}\,(0)}_{q,\,{\rm NS}}\left(z,E R, E R\right)$ from
  $E^2$ to $E^2 R^2$, and thus obtain the final form of the evolution
  equation for the NS channel
  \begin{equation}\label{eq:NS-RGE-final}
 \frac{ d D_{\rm  NS}^{\rm jet}(z,\mu, E R)}{d\ln\mu^2}
  =\frac{\alpha_s(\mu^2)}{2\pi}
  \left(\hat{P}_{qq}^{(0)}(z)+\frac{\alpha_s(\mu^2)}{2\pi}
    \hat{P}_{qq}^{(1),\,{\rm V}}(z) -\frac{\alpha_s(E^2R^2)}{2\pi}
    \delta\hat{P}_{qq}^{(1)}\right)\otimes D_{\rm  NS}^{\rm jet}(z,\mu, E R)\,.
  \end{equation}
Analogous considerations apply to the other flavour channels.

We conclude with a final remark. In Eqs.~\eqref{eq:t},~\eqref{eq:Pq}
we have expanded out, compared to the notation in Eq.~(2.1) of
Ref.~\cite{vanBeekveld:2023lsa}, the dependence on the momentum
fraction $x$ from the coupling constant. One could avoid this
expansion and instead directly embed this dependence into the
structure of the evolution equation. In this case it is convenient to
write an evolution equation in the lower bound of the angular
evolution range rather than in the upper one, leading to
\begin{align}\label{eq:NS-RGE-final-2}
 \frac{ d D_{\rm  NS}^{\rm jet}(z,E, \mu)}{d\ln\mu^2}
  = & - \hat{P}_{qq}^{(0)}(z) \otimes \left(\frac{\alpha_s(z^2\mu^2)}{(2\pi)} \,D_{\rm  NS}^{\rm jet}(z,E, \mu)\right)\notag\\
& -    \hat{P}_{qq}^{(1),\,{\rm NS}}(z)\otimes \left(\frac{\alpha^2_s(z^2\mu^2)}{(2\pi)^2} \,D_{\rm  NS}^{\rm jet}(z,E, \mu)\right)\,,
\end{align}
where $\hat{P}_{qq}^{(1),\,{\rm NS}}(z)$ is defined in
Eq.~\eqref{eq:PNS}.
This equation can be solved, with the same boundary condition given in
Eq.~\eqref{eq:DNSR-boundary} evaluated at $\mu=E\,R$, by integrating
between $\mu=E\,R$ and $\mu = E$. We notice that, in this case, the
jet radius $R$ does not appear in the running coupling, but it is
encoded in the structure of the differential equation. We have
compared the NLL solutions to
Eqs.~\eqref{eq:NS-RGE-final},~\eqref{eq:NS-RGE-final-2}, finding
complete agreement.

\bibliographystyle{JHEP}
\bibliography{triple-collinear}
\end{document}